\documentclass[12pt, letterpaper]{article}

\usepackage{amsmath, amsfonts, color}
\usepackage{graphicx}
\usepackage{natbib}
\usepackage{enumerate}
\usepackage{setspace}
\usepackage{afterpage}
\usepackage{caption}
\usepackage{subcaption}
\usepackage{ulem}

\newcommand{\bmX}{\mbox{\boldmath $X$}}
\newcommand{\bmt}{\mbox{\boldmath $t$}}

\newcommand{\bmy}{\mbox{\boldmath $y$}}

\newcommand{\Prob}{\mbox{Pr}}
\newcommand{\mblog}{\mbox{log}}
\newcommand{\mblogit}{\mbox{logit}}

\newcommand{\bmtheta}{\boldsymbol{\theta}}

\newcommand{\hidden}{hidden }

\begin{document}

\begingroup  
  \centering
  \LARGE Predictive Modeling of Cholera Outbreaks in Bangladesh\\[1.5em]
\vspace{-10pt}
  \large Amanda A. Koepke$^{1}$, 
  Ira M. Longini, Jr.$^{2}$, M. Elizabeth Halloran$^{1,3}$, \\
  Jon Wakefield$^{3,4}$, and  Vladimir N. Minin$^{4,5,*}$\\
\vspace{5pt}
  \footnotesize{$^{1}$Fred Hutchinson Cancer Research Center, Seattle, Washington,
  U.S.A.} \\
  \footnotesize{$^{2}$Department of Biostatistics and Emerging Pathogens Institute,
  University of Florida,
  Gainesville, Florida, U.S.A.}\\
  \footnotesize{$^{3}$Department of Biostatistics, University of Washington,
  Seattle, Washington, U.S.A.} \\
  \footnotesize{$^{4}$Department of Statistics, University of Washington, Seattle, Washington, U.S.A.}\\
  \footnotesize{$^{5}$Department of Biology, University of Washington, Seattle,
  Washington, U.S.A. } \\
  \footnotesize{ $*$email: vminin@uw.edu}\par
\endgroup

\date{}

\begin{abstract}
  Despite seasonal cholera outbreaks in Bangladesh, little is known
  about the relationship between environmental conditions and cholera
  cases. We seek to develop a predictive model for cholera outbreaks
  in Bangladesh based on environmental predictors. To do this, we
  estimate the contribution of environmental variables, such as water
  depth and water temperature, to cholera outbreaks in the context of
  a disease transmission model. We implement a method which
  simultaneously accounts for disease dynamics and environmental
  variables in a Susceptible-Infected-Recovered-Susceptible (SIRS)
  model. The entire system is treated as a continuous-time hidden
  Markov model, where the hidden Markov states are the numbers of
  people who are susceptible, infected, or recovered at each time
  point, and the observed states are the numbers of cholera cases
  reported. We use a Bayesian framework to fit this hidden SIRS model,
  implementing particle Markov chain Monte Carlo methods to sample
  from the posterior distribution of the environmental and
  transmission parameters given the observed data. We test this method
  using both simulation and data from Mathbaria,
  Bangladesh. Parameter estimates are used to make short-term
  predictions that capture the formation and decline of epidemic
  peaks. We demonstrate that our model can successfully predict an
  increase in the number of infected individuals in the population
  weeks before the observed number of cholera cases increases, which
  could allow for early notification of an epidemic and timely
  allocation of resources.
\end{abstract}


\thispagestyle{empty}
\clearpage
\setcounter{page}{1} 

\section{Introduction}
\label{s:intro}

In Bangladesh, cholera is an endemic disease that demonstrates
seasonal outbreaks \citep{phase1,koelle2004,koelle2005,33yr}. The
burden of cholera is high in that country, with an estimated 352,000
cases and 3,500 to 7,000 deaths annually
\citep{cholerainvestmentcasebangladesh12}. We seek to understand the
dynamics of cholera and to develop a model that will be able to
predict outbreaks several weeks in advance. If the timing and size of
a seasonal epidemic could be predicted reliably, vaccines and other
resources could be allocated effectively to curb the impact of the
disease.

Specifically, we want to understand how the disease dynamics are
related to environmental covariates. It is currently not known what
triggers the seasonal cholera outbreaks in Bangladesh, but it has been
shown that \textit{Vibrio cholerae}, the causative bacterial agent of
cholera, can be detected in the environment year round
\citep{colwell1,colwell2}. Environmental forces are thought to
contribute to the spread of cholera, evident from the many cholera
disease dynamics models that incorporate the role of the aquatic
environment on cholera transmission through an environmental reservoir
effect \citep{codeco,tien}. One hypothesis is that proliferation of
\textit{V. cholerae} in the environment triggers the seasonal
epidemic, feedback from infected individuals drives the epidemic, and
then cholera outbreaks wane, either due to an exhaustion of the
susceptibles or due to the deteriorating ecological conditions for
propagation of \textit{V. cholerae} in the environment. We probe this
hypothesis using cholera incidence data and ecological data collected
from multiple thanas (administrative subdistricts with a police
station) in rural Bangladesh over sixteen years. There have been three
phases of data collection so far, each lasting approximately three
years and being separated by gaps of a few years; the current
collection phase is ongoing.  For a subset of these data,
\cite{phase1} used Poisson regression to study the association between
lagged predictors from a particular water body to cholera cases in
that thana. This resulted in different lags and different significant
covariates across multiple water bodies and thanas. Thus, it was hard
to derive a cohesive model for predicting cholera outbreaks from the
environmental covariates. Also, there is no easy way to account for
disease dynamics in this Poisson regression framework. We want to
measure the effect of the environmental covariates while accounting
for disease dynamics via mechanistic models of disease
transmission. Moreover, we want to see if we can make reliable
short-term predictions with our model --- a task that was not
attempted by \cite{phase1}.

Mechanistic infectious disease models use scientific understanding of
the transmission process to develop dynamical systems that describe
the evolution of the process \citep{breto}. Realistic models of
disease transmission incorporate non-linear dynamics \citep{he}, which
leads to difficulties with statistical inference under these models,
specifically in the tractability of the likelihood.
\citet{keelingross} demonstrate some of these difficulties; they use
an exact stochastic continuous-time, discrete-state model which
evolves Markov processes using the deterministic Kolmogorov forward
equations to express the probabilities of being in all possible
states. However, that method only works for small populations due to
computational limitations. To overcome this intractability,
\cite{TSIR2000} develop a time-series Susceptible-Infected-Recovered
(SIR) model which extends mechanistic models of disease dynamics to
larger populations. A similar development is the auto-Poisson model of
\cite{hhh4}. To facilitate tractability of the likelihood, both of the
above approaches make simplifying assumptions that are difficult to
test.  Moreover, these discrete-time approaches work only for evenly
spaced data or require aggregating the data into evenly spaced
intervals. \cite{cauchemez2008likelihood} develop a different,
continuous-time, approach to analyze epidemiological time-series data,
but assume the transmission parameter and number of susceptibles
remain relatively constant within an observation period. Our current
understanding of cholera disease dynamics leads us to think that this
assumption is not appropriate for modeling endemic cholera with
seasonal outbreaks.

To implement a mechanistic approach without these approximations, both
maximum likelihood and Bayesian methods can be used. Maximum
likelihood based statistical inference techniques use Monte Carlo
methods to allow maximization of the likelihood without explicitly
evaluating it \citep{he,breto,ionides2006,pompmalaria}.
\cite{ionides2006} use this methodology to study how large scale
climate fluctuations influence cholera transmission in
Bangladesh. \cite{pompmalaria} use this framework to study malaria
transmission in India. They are able to incorporate a rainfall
covariate into their model and study how climate fluctuations
influence disease incidence when one controls for disease dynamics,
such as waning immunity. Under a Bayesian approach, particle filter
Markov chain Monte Carlo (MCMC) methods have been developed which
require only an unbiased estimate of the likelihood
\citep{pmcmclong}. \cite{ras} use this particle MCMC methodology to
simultaneously estimate the epidemiological parameters of a SIR model
and past disease dynamics from time series data and gene
genealogies. Using Google flu trends data \citep{googleflu},
\cite{dukic} implement a particle filtering algorithm which
sequentially estimates the odds of a pandemic. Notably, \cite{dukic}
concentrate on predicting influenza activity. Similarly, here we
develop a model-based predictive framework for seasonal cholera
epidemics in Bangladesh.

In this paper, we use sequential Monte Carlo methods in a Bayesian
framework. Specifically, we develop a hidden
Susceptible-Infected-Recovered-Susceptible (SIRS) model for cholera
transmission in Bangladesh, incorporating environmental covariates.
We use a particle MCMC method to sample from the posterior
distribution of the environmental and transmission parameters given
the observed data, as described by \cite{pmcmclong}. Further, we
predict future behavior of the epidemic within our Bayesian
framework. Cholera transmission dynamics in our model are described by
a continuous-time, rather than a discrete-time, Markov process to
easily incorporate data with irregular observation times. Also, the
continuous-time framework allows for greater parameter
interpretability and comparability to models based on deterministic
differential equations. We test our Bayesian inference procedure using
simulated cholera data, generated from a model with a time-varying
environmental covariate. We then analyze cholera data from Mathbaria,
Bangladesh, similar to the data studied by \cite{phase1}. Parameter
estimates indicate that most of the transmission is coming from
environmental sources. We test the ability of our model to make
short-term predictions during different time intervals in the data
observation period and find that the pattern of predictive
distribution dynamics matches the pattern of changes in the reported
number of cases. Moreover, we find that the predictive distribution of
the hidden states, specifically the unobserved number of infected
individuals, clearly pinpoints the beginning of an epidemic
approximately two to three weeks in advance, making our methodology
potentially useful during cholera surveillance in Bangladesh.

\section{SIRS model with environmental predictors}\label{sec:SIRS}

\begin{figure}[h!]
\vspace{-10pt}
  \begin{centering}
    \includegraphics[width=.9\linewidth]{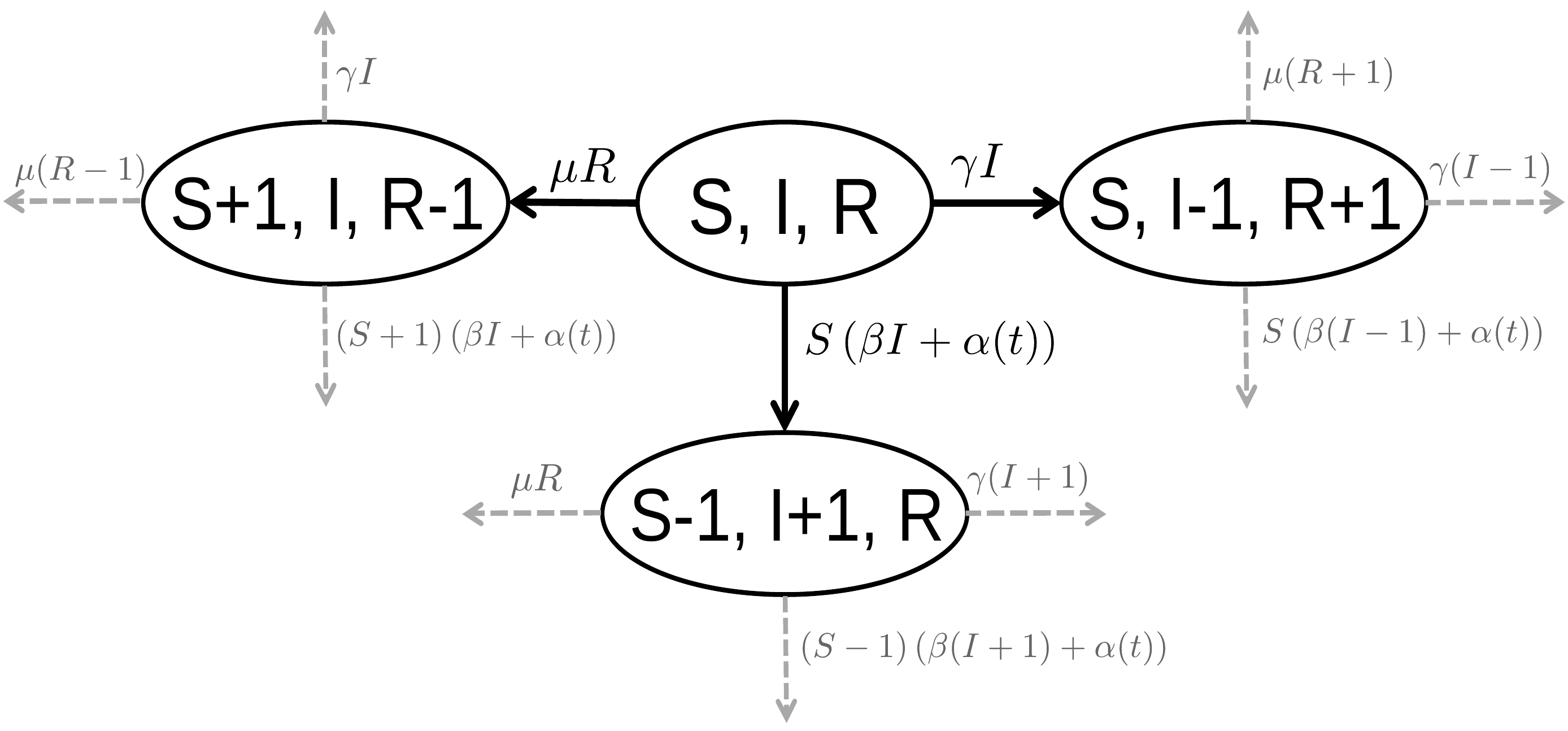}
\vspace{-10pt}
\caption{State transitions for
  Susceptible-Infected-Recovered-Susceptible (SIRS) model for
  cholera. $S$, $I$, and $R$ denote the numbers of susceptible,
  infected, and recovered individuals. From the current state
  $(S,I,R)$, the system can transition to one of three new
  states. These new states correspond to a susceptible becoming
  infected, an infected recovering from infection, or a recovered
  individual losing immunity to infection and becoming susceptible.
  The parameter $\beta$ is the infectious contact rate, $\alpha(t)$ is
  the time-varying environmental force of infection, $\gamma$ is the
  recovery rate, and $\mu$ is the rate at which immunity is lost. }
    \label{SIRS}
  \end{centering}
\vspace{-10pt}
\end{figure}

We consider a compartmental model of disease transmission
\citep{andersonmay, keeling}, where the population is divided into
three disease states, or compartments: susceptible, infected, and
recovered. We model a continuous process observed at discrete time
points. The vector $\bmX_t=(S_t,I_t,R_t)$ contains the numbers of
susceptible, infected, and recovered individuals at time $t$, and we
consider a closed population of size $N$ such that $N=S_t+I_t+R_t$ for
all $t$. Individuals move between the compartments with different
rates; for cholera transmission we consider the transition rates shown
in Figure \ref{SIRS}. In this framework, a susceptible individual's
rate of infection is proportional to the number of infected people and
the covariates that serve as proxy for the amount of
\textit{V. cholerae} in the environment. Thus, the hazard rate of
infection, also called the force of infection, is $\beta I_t +
\alpha(t) $ for each time $t$, where $\beta$ represents the infectious
contact rate between infected individuals and susceptible individuals
and $\alpha(t)$ represents the time-varying environmental force of
infection. Possible mechanisms for infectious contact include direct
person-to-person transmission of cholera and consumption of water that
has been contaminated by infected individuals. Infected individuals
recover from infection at a rate $\gamma$, where $1/\gamma$ is the
average length of the infectious period. Once the infected individual
has recovered from infection, they move to the recovered
compartment. Recovered individuals develop a temporary immunity to the
disease after infection. They move from the recovered compartment to
the susceptible compartment with rate $\mu$, where $1/\mu$ is the
average length of immunity. Similar to \cite{codeco} and
\cite{koelle2004}, birth and death are incorporated into the system
indirectly through the waning of immunity; thus, instead of
representing natural loss of immunity only, $\mu$ also represents the
loss of immunity through the death of recovered individuals and birth
of new susceptible individuals.

We model $\bmX_{t}$ as an inhomogeneous Markov process
\citep{stochintro} with infinitesimal rates

\begin{eqnarray}\label{infgen}
  \lambda_{(S,I,R),(S',I',R')}(t) = \left\{ \begin{array}{rl}
      \left( \beta I + \alpha(t) \right) S &\mbox{ if $S'=S-1$, $I' = I+1$, $R'=R$}, \\
      \gamma I &\mbox{ if $S'=S$, $I' = I-1$, $R'=R+1$}, \\
      \mu R &\mbox{ if $S'=S+1$, $I' = I$, $R'=R-1$}, \\
      0 &\mbox{ otherwise,}
    \end{array} \right.
\end{eqnarray}
where $\bmX=(S,I,R)$ is the current state and $\bmX'=(S',I',R')$ is a
new state. Because $R_t=N-S_t-I_t$, we keep track of only susceptible
and infected individuals, $S_{t}$ and $I_{t}$.

This type of compartmental model is similar to other cholera models in
the literature. The time-series SIRS model of \cite{koelle2004} also
includes the effects of both intrinsic factors (disease dynamics) and
extrinsic factors (environment) on transmission. \cite{king2008}
examine both a regular SIRS model and a two-path model to include
asymptomatic infections, and use a time-varying transmission term
that incorporates transmission via the environmental reservoir and
direct person-to-person transmission, but does not allow for feedback
from infected individuals into the environmental reservoir. The SIWR
model of \cite{tien} and \cite{eisen2013} allows for infections from
both a water compartment (W) and direct transmission and considers the
feedback created by infected individuals contaminating the water. To
allow for the possibility of asymptomatic individuals, \cite{control}
use a model with a compartment for asymptomatic infections; that model
only considers direct transmission. \cite{codeco} uses an SIR model
with no direct person-to-person transmission; infected individuals
excrete directly into the environment and susceptible individuals are
infected from exposure to contaminated water. Our SIRS model is not
identical to any of the above models, but it borrows from them two
important features: explicit modeling of disease transmission from
either direct person-to-person transmission of cholera or consumption
of water that has been contaminated by infected individuals and a
time-varying environmental force of infection.

\section{Hidden SIRS model}\label{sec:HMM}

While the underlying dynamics of the disease are described by
$\bmX_t$, these states are not directly observed. The number $y_t$ of
infected individuals observed at each time point $t$ is only a random
fraction of the number of infected individuals. This fraction depends
on both the number of infected individuals that are symptomatic and
the fraction of symptomatic infected individuals that seek treatment
and get reported (the reporting rate). Thus, $y_{t_i}$, the number of
observed infections at time $t_i$ for observation $i \in
\{0,1,...,n\}$, has a binomial distribution with size $I_{t_i}$, the
number of infected individuals at time $t_i$, and success probability
$\rho$, the probability of infected individuals seeking treatment, so
\begin{eqnarray}\label{ydist}
  \mbox{Pr}(y_{t_i}|I_{t_i},\rho) = {I_{t_i} \choose y_{t_i}} \rho^{y_{t_i}}(1-\rho)^{I_{t_i}-y_{t_i}}. 
\end{eqnarray} 
Given $\bmX_{t_i}$, $y_{t_i}$ is independent of the other
observations and other hidden states. 

We use a Bayesian framework to estimate the parameters of the hidden
SIRS model, where the unobserved states $\bmX_t$ are governed by the
infinitesimal rates in Equation (\ref{infgen}). The parameters that we
want to estimate are $\beta$, $\gamma$, $\mu$, $\rho$, and the $k+1$
parameters that will be incorporated into $\alpha(t)$, the
time-varying environmental force of infection. We assume
$\alpha(t)=\mbox{exp} \left (
  \alpha_0+\alpha_1C_1(t)+\dots+\alpha_kC_k(t)\right ) $, where
$C_1(t),\ldots,C_k(t)$ denote the $k$ time-varying environmental
covariates.

We assume independent Poisson initial distributions for $S_{t_0}$ and
$I_{t_0}$, with means $\phi_{S}$ and $\phi_{I}$. Thus
$$\Prob(\bmX_{t_0}|\phi_{S},\phi_{I}) = \Prob(S_{t_0}|\phi_{S}) \times
\Prob(I_{t_0}|\phi_{I}) =
\frac{\phi_{S}^{S_{t_0}}\mbox{exp}(-\phi_{S})}{S_{t_0}!} \times
\frac{\phi_{I}^{I_{t_0}}\mbox{exp}(-\phi_{I})}{I_{t_0}!}.$$ Parameters
that are constrained to be greater than zero, such as $\beta$,
$\gamma$, $\mu$, $\phi_{S}$, and $\phi_{I}$, are transformed to the
log scale. A logit transformation is used for the probability
$\rho$. We assume independent normal prior distributions on all of the
transformed parameters, incorporating biological information into the
priors where possible.

We are interested in the posterior distribution
$\mbox{Pr}(\bmtheta|\bmy) \propto
\mbox{Pr}(\bmy|\bmtheta)\mbox{Pr}(\bmtheta)$, where
$\bmy=(y_{t_0},\ldots,y_{t_n})$, $\bmtheta = \left(\mblog(\beta), \mblog(\gamma), \mblog(\mu),
  \mbox{logit}(\rho), \alpha_0,\dots,\alpha_k,\mblog(\phi_{S}),\mblog(\phi_{I}) \right)$, and
\begin{eqnarray*}
  \mbox{Pr}(\mathbf{y}|\bmtheta)
  &=& \sum_{\tiny{\bmX}} \left ( \prod_{i=0}^n \mbox{Pr}(y_{t_i}|I_{t_i},\rho) \left[ \Prob(\bmX_{t_0}|\phi_{S},\phi_{I}) \prod_{i=1}^n p(\bmX_{t_i}|\bmX_{t_{i-1}},\bmtheta)\right] \right ).
\end{eqnarray*}
Here $p(\bmX_{t_i}|\bmX_{t_{i-1}},\bmtheta)$ for $i=1,\ldots,n$ are
the transition probabilities of the continuous-time Markov chain
(CTMC). However, this likelihood is intractable; there is no practical
method to compute the finite time transition probabilities of the SIRS
CTMC because the size of the state space of $\bmX_t$ grows on the 
order of $N^2$. For the same reason, summing over $\bmX$ with the
forward-backward algorithm \citep{baum1970} is not feasible. To use
Bayesian inference despite this likelihood intractability, we turn to
a particle marginal Metropolis-Hastings (PMMH) algorithm.

\section{Particle filter MCMC}

\subsection{Overview}

The PMMH algorithm, introduced by \cite{beaumontpmcmc} and studied in
\cite{androberts} and \cite{pmcmclong}, constructs a Markov chain that
targets the joint posterior distribution
$\pi(\bmtheta,\bmX|\mathbf{y})$, where $\bmX$ is a set of auxiliary or
\hidden variables, and requires only an unbiased estimate of the
likelihood. To construct this likelihood estimate, we use an SMC
algorithm, also known as a bootstrap particle filter
\citep{smcbook}. The SMC algorithm sequentially estimates the
likelihood using weighted particles; it requires the ability to
propagate the unobserved data, $\bmX_t$, forward in time and the
calculation of the probability of the observed data given the
simulated unobserved data. For the hidden SIRS model,
$y_{t_i}|\bmX_{t_i}=(S_{t_i},I_{t_i},R_{t_i}),\rho \sim
\mbox{Binomial}(I_{t_i},\rho)$, where $\rho$ depends on the number of
symptomatic infected individuals that seek treatment, as described in
Section \ref{sec:HMM}. Thus the probability of the observed data given
the simulated unobserved data is given by Equation (\ref{ydist}).  To
propagate the \hidden variables forward in time, we first simulate
initial states $\bmX_{t_0}=\left(S_{t_0},I_{t_0} \right)$ from Poisson
distributions with means $\phi_{S}$ and $\phi_{I}$. We then use
properties of CTMCs to simulate the trajectories of the unobserved
states.

Thus, the PMMH algorithm has two parts: an SMC algorithm, which is
used to estimate the marginal likelihood of the data given a
particular set of parameters, $\bmtheta$, and a Metropolis-Hastings
step \citep{metropolis,hastings}, which uses the estimated likelihood
in the acceptance ratio. At each step, a new $\bmtheta^*$ is proposed
from the proposal distribution $q(\cdot|\bmtheta)$. An SMC algorithm
is used to generate and weight $K$ particle trajectories corresponding
to the \hidden state processes using the proposed parameter set
$\bmtheta^*$. A proposed
$\bmX^*_{\bmt_{0:n}}=\left(\bmX^*_{t_0},\ldots,\bmX^*_{t_n}\right)$
trajectory is sampled from the $K$ particle trajectories based on the
final particle weights of the SMC algorithm. The marginal likelihood
is estimated by summing the weights of the SMC algorithm, and the
proposed $\bmtheta^*$ and $\bmX^*_{\bmt_{0:n}}$ are accepted with
probability equal to the familiar Metropolis-Hastings acceptance
ratio.

To propagate the unobserved $\bmX_t=(S_t,I_t,R_t)$ forward in
time, we simulate from a cholera transmission model with a time-varying
environmental force of infection. CTMCs
which incorporate time-varying transition rates are inhomogeneous. The
details of the discretely-observed inhomogeneous CTMC simulations are now described.

\subsection{Simulating inhomogeneous SIRS using
  tau-leaping}\label{sec:inhomoTL}

Gillespie developed two methods for exact stochastic simulation of
trajectories with constant rates: the direct method \citep{gill1977}
and the first reaction method \citep{gill1976}. Details of these
methods are given in Appendix A. The exact algorithms work for small
populations, but for large state spaces these methods require a
prohibitively long computing time. This is a common problem in the
chemical kinetics literature, where an approximate method called the
tau-leaping algorithm originated \citep{gill2001,gillmodTL}. This
method simulates CTMCs by jumping over a small amount of time $\tau$
and approximating the number of events that happen in this time using
a series of Poisson distributions. As $\tau$ approaches zero, this
approximation theoretically approaches the exact algorithm. The value
of $\tau$ must be chosen such that the rates remain roughly constant
over the period of time; this is referred to as the ``leap
condition''.

Specifically, for our simulation, using the methods outlined in
\cite{gillmodTL}, we define the rate functions $h_1(\bmX_t)= \left(
  \beta I_t+\alpha(t) \right) S_t$, $h_2(\bmX_t) = \gamma I_t$, and
$h_3(\bmX_t) = \mu R_t$, corresponding to the infinitesimal rates of
the CTMC. Then $k_1 \sim \mbox{Poisson}(h_1(\bmX_t) \tau)$ represents
the number of infections in time $[t,t+\tau)$, $k_2 \sim
\mbox{Poisson}(h_2(\bmX_t) \tau)$ represents the number of recoveries
in time $[t,t+\tau)$, and $k_3 \sim \mbox{Poisson}(h_3(\bmX_t) \tau)$
represents the number of people that become susceptible to infection
in time $[t,t+\tau)$. We make the assumption that the time-varying
force of infection, $\alpha(t)$, remains constant each day. We define
daily time intervals $A_i := [i,i+1)$ for $i \in
\{t_0,t_0+1,\ldots,t_n-1\}$, and $\alpha(t)=\alpha_{A_i}$ for $t \in
A_i$. Using $\tau=1$ day, our rates now remain constant within each
tau jump. To see if this value for $\tau$ is reasonable, we perform a
simulation study; see Appendix A for details.

\subsection{Metropolis-Hastings proposal for model parameters}

Our implementation of the PMMH algorithm starts with a preliminary
run, which consists of a burn-in run plus a secondary run, both using
independent normal random walk proposal distributions for the
parameters. From the secondary run, we calculate the approximate
posterior covariance of the parameters and use it to construct the
covariance of the multivariate normal random walk proposal
distribution in the final run of the PMMH algorithm. In all runs,
parameters are proposed and updated jointly.

\subsection{Prediction}

One of the main goals of this analysis is to be able to predict
cholera outbreaks in advance using environmental predictors.  To
assess the predictive ability of our model, we estimate the parameters
of the model using a training set of data and then predict future
behavior of the epidemic process.  We examine the posterior predictive
distributions of cholera counts by simulating data forward in time
under the time-varying SIRS model using the accepted parameter values
explored by the particle MCMC algorithm and the accepted values of the
\hidden states $S_T$ and $I_T$ at the final observation time, $t=T$,
of the training data. These \hidden states are sampled in the PMMH
algorithm by sampling the last set of particles using the last set of
weights \citep{pmcmclong}.  Under each set of parameters, we generate
possible future \hidden states and observed data, and we compare the
posterior predictive distribution of observed cholera cases to the
test data. In the analyses below, the PMMH output is always thinned to
500 iterations for prediction purposes by saving only every $k$th
iteration, where $k$ depends on the total number of iterations.

\section{Simulation results}\label{sec:inhomoPMMH}

\begin{figure}[h!]
\centering
    \includegraphics[width=1\textwidth]{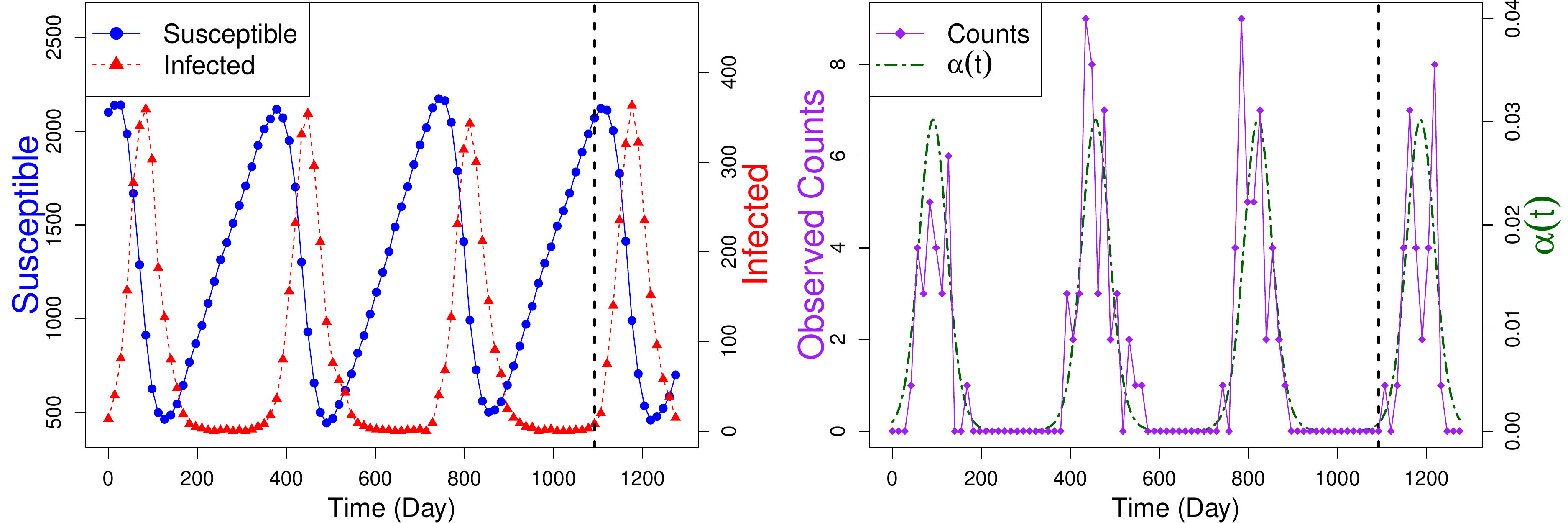}
\label{inhomosim}
  \vspace{-20pt}
  \caption{Plots of simulated \hidden states (counts of susceptible,
    $S_t$, and infected, $I_t$, individuals) and the observed data
    (number of observed infections $=y_t \sim \mbox{Binomial} (I_t,\rho)$ ) plotted over time, $t$. Simulation with seasonally
    varying $\alpha(t)$ generates data with seasonal epidemic
    peaks. The dashed vertical black line represents the first cut off
    between the training sets and the test data. Data before the line
    are used to estimate parameters, and we use those estimates to
    predict the data after the line. Other data cut offs are shown in
    Figure \ref{predsimpred}}
  \label{simdata}
  \vspace{-10pt}
\end{figure}

To test the PMMH algorithm on simulated infectious disease data, we
generate data from a hidden SIRS model with a time-varying
environmental force of infection. We then use our Bayesian framework
to estimate the parameters of the simulated model and compare the
posterior distributions of the parameters with the true values. To
simulate endemic cholera where many people have been previously
infected, we start with a population size of $N=10000$ and assume
independent Poisson initial distributions for $S_{t_0}$ and $I_{t_0}$,
with means $\phi_{S}=2100$ and $\phi_{I}=15$. The other parameters are
set at $\beta =1.25\times 10^{-5}$, $\gamma =0.1$, and $\mu =
0.0009$. All rates are measured in the number of events per day. The
average length of the infectious period, $1/\gamma$, is set to be 10
days, and the average length of immunity, $1/\mu$, is set to be about
3 years. Parameter values are chosen such that the simulated data are
similar to the data collected from Mathbaria, Bangladesh. We use the
daily time intervals $A_i := [i,i+1)$ for $i \in
\{t_0,t_0+1,\ldots,t_n-1\}$, as in Section \ref{sec:inhomoTL}, and
define $\alpha(t)=\alpha_{A_i}$ for $t \in A_i$ where
$\alpha_{A_i}=\mbox{exp}\left[\alpha_0 + \alpha_1
  \mbox{sin}\left(2\pi i /365\right)\right].$ The intercept
$\alpha_0$ and the amplitude $\alpha_1$ are parameters to be
estimated. The frequency of the sine function is set to mimic the
annual peak seen in the environmental data collected from
Bangladesh. For the simulations we set $\alpha_0 =-7$ and $\alpha_1
=3.5$. Using the modified Gillespie algorithm described in Appendix A,
we simulate the $(S_t, I_t)$ chain given in the left plot of
Figure~\ref{simdata}. The observed number of infections $y_t \sim
\mbox{Binomial} (I_t,\rho)$, where $\rho=0.015$ and is treated as an
unknown parameter.

We simulate three years of training data because this is approximately
how long the data collection phases last in our data from Bangladesh
\citep{phase1}.  Therefore we do not attempt to estimate the loss of
immunity rate $\mu$ since it is on the scale of three years. Also,
there is not enough information in the data to estimate the means of
the Poisson initial distributions, $\phi_S$ and $\phi_I$, since
estimation of these parameters is only informed by the very beginning
of the observed data. We set these parameters to different values and
compare parameter estimation and prediction between models with
parameter assumptions which differ from the truth. We also assume that
we know the population size, $N=10000$.

We assume normal prior distributions on all of the 
parameters, with means and standard deviations chosen such that the
mass of each prior distribution is not centered at true value of the
parameter in this simulation setting. We use relatively uninformative,
diffuse priors for $\mblog(\beta)$, $\alpha_0$, and $\alpha_1$,
centered at $\mblog(1.25 \times 10^{-4})$, $-8$, and 0, respectively,
and with standard deviations of 5. The prior distribution for
$\mblogit(\rho)$ is centered at $\mbox{logit}(0.03)$ and has a
standard deviation of 2. For $\mblog(\gamma)$, the prior is centered
at $\mblog(0.1)$ with a relatively small standard deviation of $0.09$,
since this value is well studied for cholera. Thus, \textit{a priori} $1/\gamma$ 
falls between 8.4 to 11.9 days with probability 0.95.

Using these data, the PMMH algorithm starts with a burn-in run of
10000 iterations, a secondary run of 10000 iterations, and a final run
of 50000 iterations. To thin the chains, we save only every 10th
iteration. We use $K=100$ particles in the SMC algorithm. We compare
results from models with different assumptions on the values of
$\phi_S$ and $\phi_I$: assumed $\phi_S/N$ and $\phi_I/N$ are above the
true values (0.31 and 0.003), at the true values (0.21 and 0.0015),
below the true values (0.11 and 0.00075), or further below the true
values (0.055 and 0.000375). Marginal posterior distributions for the
parameters of the SIRS model from the final runs of these PMMH
algorithms are in Appendix B. The posterior distributions are similar,
regardless of assumed values for $\phi_S$ and $\phi_I$. Trace plots,
auto-correlation plots, bivariate scatterplots, and effective sample
sizes for the posterior samples under the situation in which the true
values of $\phi_S$ and $\phi_I$ were assumed are also given in
Appendix B. We report $\beta \times N$ and $\rho \times N$, since in
sensitivity analyses we found these to be robust to assumptions about
the total population size $N$. From the posterior distributions, it is
clear that the algorithm is providing good estimates of the true
parameter values, though estimates of the parameter $\rho \times N$
are slightly different than the truth, especially when $\phi_S$ and
$\phi_I$ are not set at the true values.

\subsection{Prediction results}
\begin{figure}[h!]
  \begin{centering}
\vspace{-10pt}
      \includegraphics[width=.9\linewidth]{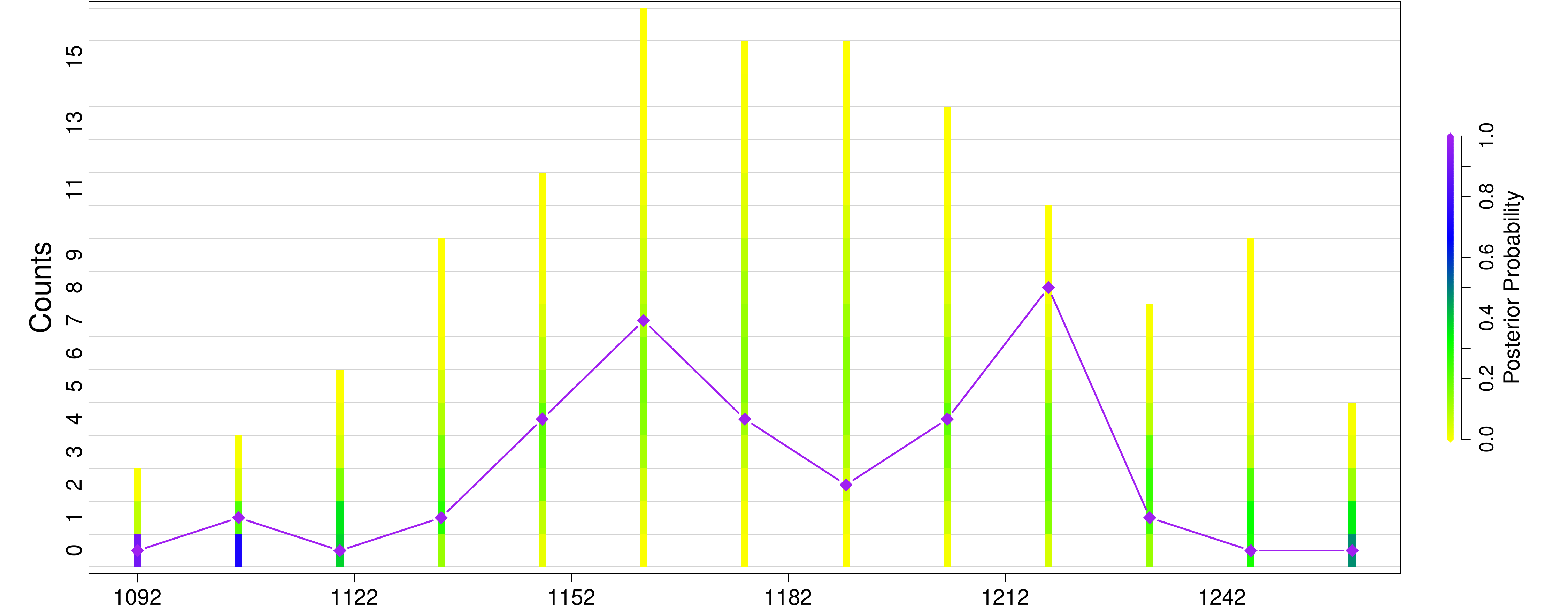}
      \includegraphics[width=.9\linewidth]{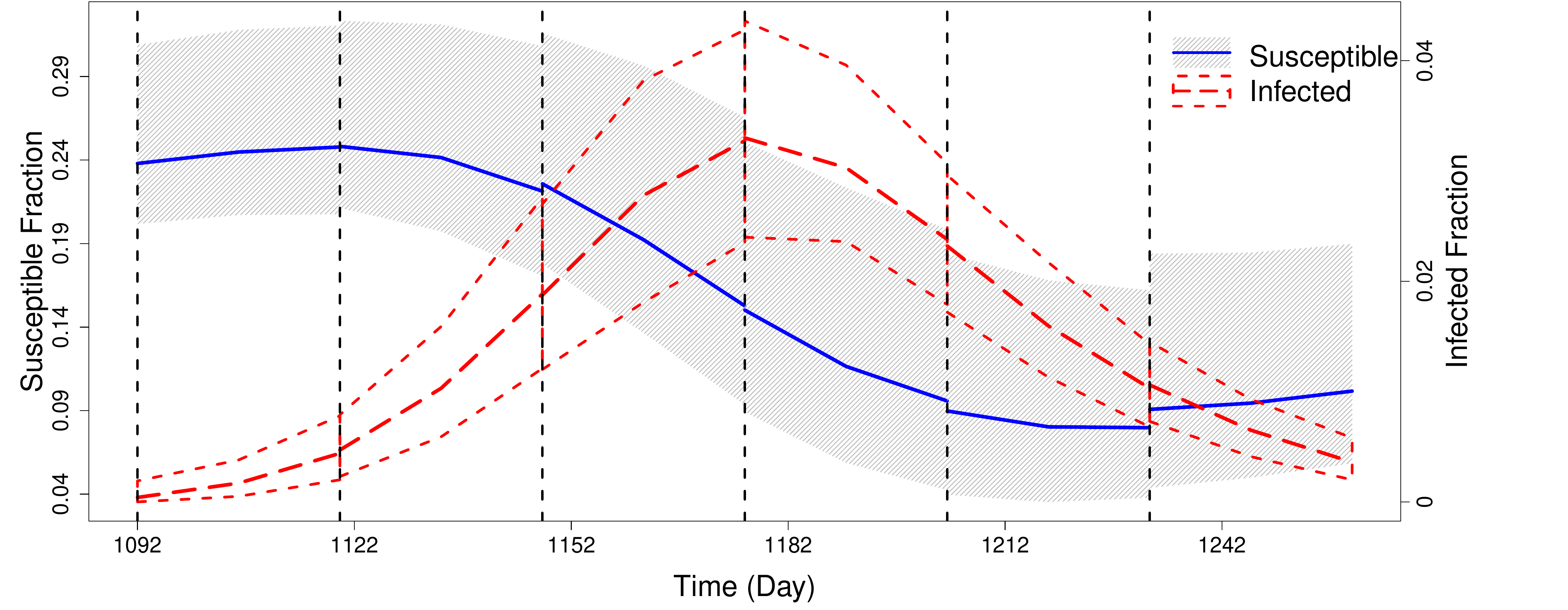}
\vspace{-10pt}
    \caption{Summary of prediction results for simulated data. We run
      PMMH algorithms on training sets of the data, which are cut off
      at each of the dashed black lines in the bottom plot. Future
      cases are then predicted until the next cut off.  The top plot
      compares the posterior probability of the predicted counts to
      the test data (diamonds connected by straight purple lines). The
      coloring of the bars is determined by the frequency of each set
      of counts in the predicted data for each time point. The bottom
      plot shows how the trajectory of the predicted hidden states
      changes over the course of the epidemic. The gray area and the
      solid line denote the 95\% quantiles and median, respectively,
      of the predictive distribution for the fraction of
      susceptibles. The short dashed lines and the long dashed line
      denote the 95\% quantiles and median, respectively, of the
      predictive distribution for the fraction of infected
      individuals.}
    \label{predsimpred}
  \end{centering}
\end{figure}

To test the predictive ability of the model, we use multiple cut off
times to separate our simulated data into staggered training sets and
test sets. The simulated observed data are shown in the right plot of
Figure \ref{simdata}. For each cut off time, parameters were drawn
from the posterior distribution based on the training data. These
parameter values were then used to simulate possible realizations of
reported infections after the training data until the next cut off, 28
days later. The distributions of these predicted reported cases are
shown in the top plot of Figure \ref{predsimpred}. The test data are
denoted by the purple diamonds, connected by straight lines to help
visualize ups and downs in the case counts. Case counts are observed
once every 14 days. On each observation day, the colored bar
represents the distribution of predicted counts for that day. As
desired, the posterior predictive distribution shifts its mass as time
progresses to follow the case counts in the test data. The plot of the
predicted \hidden states in the bottom row of Figure \ref{predsimpred}
also shows that our model is capturing the formation and decline of
the epidemic peak well, as seen in the trajectory of the predicted
fraction of infected individuals. This plot illustrates the interplay
of the hidden states of the underlying compartmental model. During an
epidemic, the fraction of susceptibles decreases while the fraction of
infected individuals quickly increases. Afterwards, the fraction of
infected individuals drops and the pool of susceptibles slowly begins
to increase as both immunity is lost and more susceptible individuals
are born.

These predictions were made under the assumption that $\phi_S$ and
$\phi_I$ are set to the true values. To test sensitivity to these
assumptions, we compare predictions made from models that assume other
values; these are shown in Appendix D. Predicted distributions are
similar for all values of $\phi_S$ and $\phi_I$.

\section{Using cholera incidence data and covariates from Mathbaria, Bangladesh}\label{sec:SIRSMath}

\begin{figure}[h!]
\vspace{-10pt}
  \begin{centering}
    \includegraphics[width=1\linewidth]{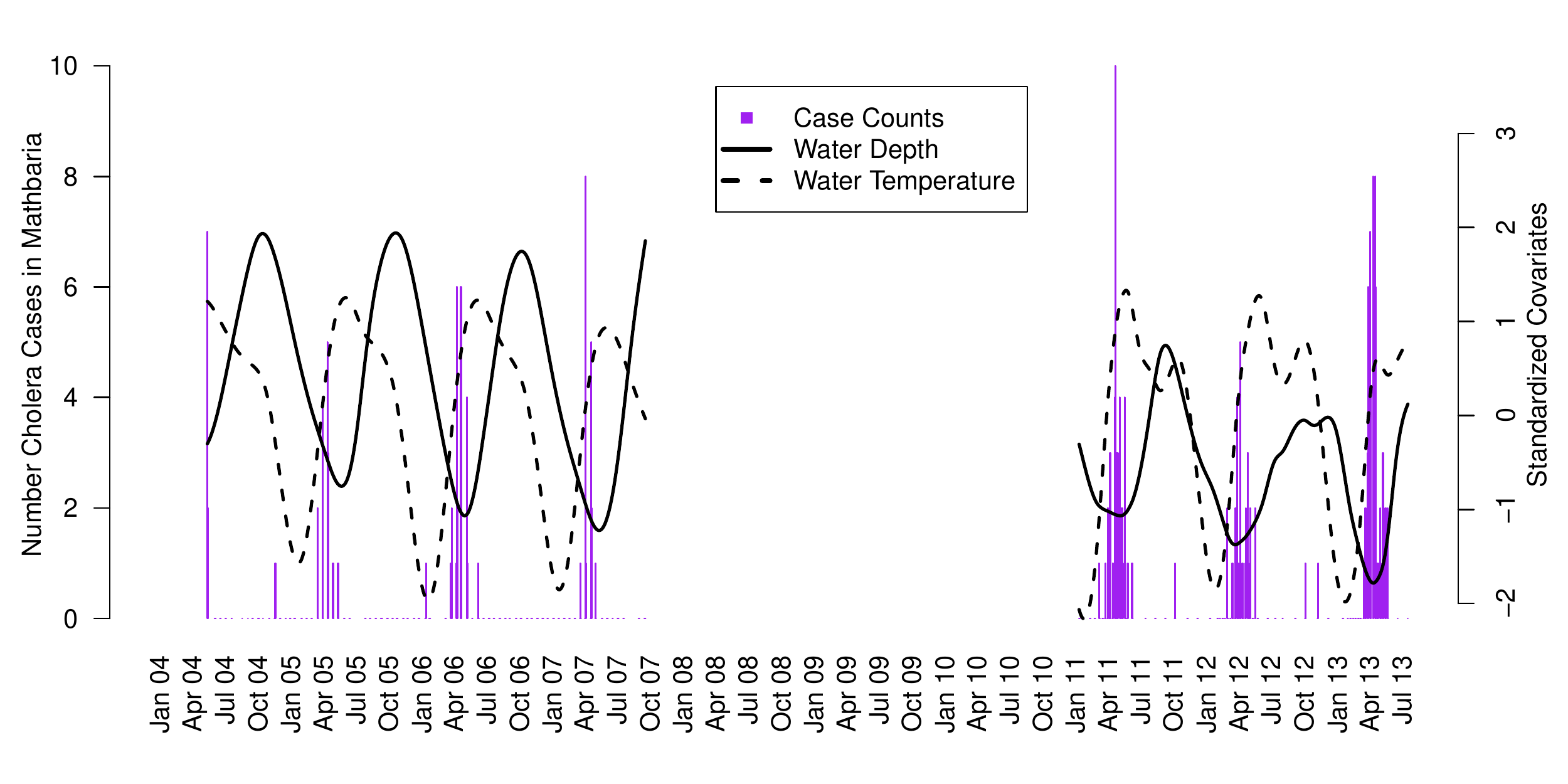}
\vspace{-30pt}
    \caption{Barplot of cholera case counts in Mathbaria, Bangladesh
      and the standardized covariate measurements over time. The
      covariates are shown with a lag of three weeks. No data were
      collected from October 2007 through November 2010. The ranges of
      the unstandardized smoothed covariates are 1.4 to 2.8 meters
      for water depth and 21.6 to 33.1$^{\circ}$C for water
      temperature.}
    \label{cholbp}
  \end{centering}
\end{figure}

\cite{phase1} found that water temperature (WT) and water depth (WD)
in some water bodies had a significant lagged relationship with
cholera incidence.  Therefore, we use these covariates and cholera
incidence data from Mathbaria, Bangladesh collected between April 2004
to September 2007 and again from October 2010 to July 2013. During
these time periods, cholera incidence data were collected over a
period of three days approximately every two weeks. Environmental data
were also collected approximately every two weeks from six water
bodies. To get a smooth summary of the covariates using data from all
water bodies, we fit a cubic spline to the covariate values. We then
slightly modify our environmental force of infection to allow for a
lagged covariate effect. Let $\kappa$ denote the length of the lag. We
consider the daily time intervals $A_{i} := [i,i+1)$ for $i \in
\{t_0,t_0+1,\ldots,t_n-1\}$ and define the environmental force of
infection $\alpha(t)=\alpha_{A_{i}}$ for $t \in A_{i}$ and $t \geq
\kappa$ where $\alpha_{A_i}=\mbox{exp}\left[\alpha_0 + \alpha_1
  C_{WD}(i-\kappa) + \alpha_2 C_{WT}(i-\kappa)\right]$. Here the
covariates are the smoothed standardized daily values $C_{WT}(i) =
(WT(i)-\overline{WT})/s_{WT}$ and $C_{WD}(i) =
(WD(i)-\overline{WD})/s_{WD}$, where $\overline{X}$ is the mean of the
measurements for all $i$ and $s_X$ is the sample standard
deviation. We consider and compare results from models assuming three
different lags: $\kappa=14$, $\kappa=18$, and $\kappa=21$. Predictions
from all three models are similar, so we report only results from the
model assuming $\kappa=21$, in order to receive the earliest warning
of upcoming epidemics; see Appendix C for details and prediction
comparisons. The smoothed, standardized, 21 day lagged covariates and
cholera incidence data are shown in Figure \ref{cholbp}.

Since there are only about six years of data, estimating the loss of
immunity rate $\mu$ is infeasible. Thus, we set $\mu = 0.0009$ so that
$1/\mu$ is 3 years \citep{sack}. Also, the population size $N$, which
quantifies the size catchment area for the medical center, is assumed
to be 10000 for computational convenience. We do not know the true
value of $N$, but 10000 is a reasonable estimate and is small enough
that simulations run quickly. We studied sensitivity to these
assumptions by setting both $\mu$ and $N$ to different values,
obtaining similar results. We also again set $\phi_S$ and $\phi_I$ to
various values and the results were insensitive. See Appendix D for
details.

In these analyses, we use relatively uninformative, diffuse normal prior
distributions on the time-varying environmental covariates $\alpha_1$
and $\alpha_2$, centered at 0 and with standard deviations of 5. The
diffuse normal prior distributions on the transformed parameter values
$\mblog(\beta)$ and $\alpha_0$ are centered at $\mblog(1.25 \times
10^{-7})$ and $-8$, respectively, with standard deviations of 5. We know
that the average infectious period for cholera, $1/\gamma$, should be
between 8 and 12 days. Thus, the transformed parameter
$\mbox{log}(\gamma)$ is given a normal prior distribution with mean
$\mblog(0.1)$ and standard deviation $0.09$ to give 0.95 prior probability of
$1/\gamma$ falling within the interval $(8,12)$. We also know that $\rho$ should be very
close to zero, since only a small proportion of cholera infections are
symptomatic and a smaller proportion will be treated at the health
complex \citep{4yr}. Thus, the transformed parameter
$\mbox{logit}(\rho)$ is given a normal prior distribution with mean
$\mbox{logit}(0.0008)$ and standard deviation equal to $2$, to give 0.95 prior probability
of $\rho$ falling within the interval $(1.6 \times 10^{-5},0.04)$.

We run the PMMH algorithm with a burn-in run of 30000
iterations, a secondary run of 20000 iterations, and a final run of
400000 iterations. We again save only every 10th iteration and use
$K=100$ particles in the SMC algorithm. Posterior medians and 95\%
Bayesian credible intervals for the parameters $\beta \times N$,
$\gamma$, $\alpha_0$, $\alpha_1$, $\alpha_2$, and $\rho \times N$
generated by the final run of the PMMH algorithm are given in Table
\ref{theMathEst}. We report $\beta \times N$ and $\rho \times N$ since we
found these parameter estimates to be robust to changes in the
population size $N$ during sensitivity analyses. For more details, see
Appendix D. The credible intervals for $\alpha_1$ and $\alpha_2$ do
not include zero, so both water depth and water temperature have a
significant relationship with the force of infection. Decreasing water
depth increases the force of infection, likely due to the higher
concentration and resulting proliferation of \textit{V. cholerae} in
the environment; increasing water temperature increases the force of
infection \citep{phase1}.

The basic reproductive number, $R_0$, is the average number of
secondary cases caused by a typical infected individual in a
completely susceptible population \citep{R0}. We report $(\beta \times
N)/\gamma$, the part of the reproductive number that is related to the
number of infected individuals in the population under our model
assumptions. Our estimate of 4.35 is fairly large; it is very similar
to the reproductive number of 5 (sd=3.3) estimated by \cite{control}
using data from Matlab, Bangladesh. However, the 95\% credible
interval is wide, with the lower end being approximately 1. Moreover,
posterior median values for $\alpha(t)$ range from 0.00003 to 0.38,
while posterior median values for $\beta I_t$ only range from 0 to
0.03, suggesting that the epidemic peaks in our model are driven
mostly by the environmental force of infection. See Appendix F for
more details. However, the infectious contact rate is not zero and is
not negligible compared to the environmental force of infection.

\begin{table}[h]
  \begin{centering}
    \caption{Posterior medians and 95\% equitailed credible intervals (CIs)
      for the parameters of the SIRS model estimated using clinical
      and environmental data sampled from Mathbaria, Bangladesh.}
    \label{theMathEst}
    \begin{tabular}{rll@{,\ }r}
      Coefficient & Estimate & \multicolumn{2}{c}{95\% CIs} 
      \tabularnewline
      \hline 
      $\beta \times N$&  0.491 & (0.103 & 0.945) \\ 
      $\gamma$ &    0.115 & (0.096 & 0.142) \\ 
      $(\beta \times N)/\gamma$  &   4.35 & (0.99 & 7.15) \\ 
      $\alpha_{0}$  &    -5.32 & (-6.63 & -4.51) \\ 
      $\alpha_{1}$  &    -1.37 & (-1.98 & -0.98) \\ 
      $\alpha_{2}$  &    2.18 & (1.8 & 2.62) \\ 
      $\rho \times N$ &    55.8 & (43.4 & 73.5) \\ 
      \hline 
    \end{tabular}
    \par\end{centering}
\end{table}

\subsection{Prediction Results}\label{sec:SIRSMathpred}

\begin{figure}[h!]
  \begin{centering}
    \includegraphics[width=1.1\linewidth]{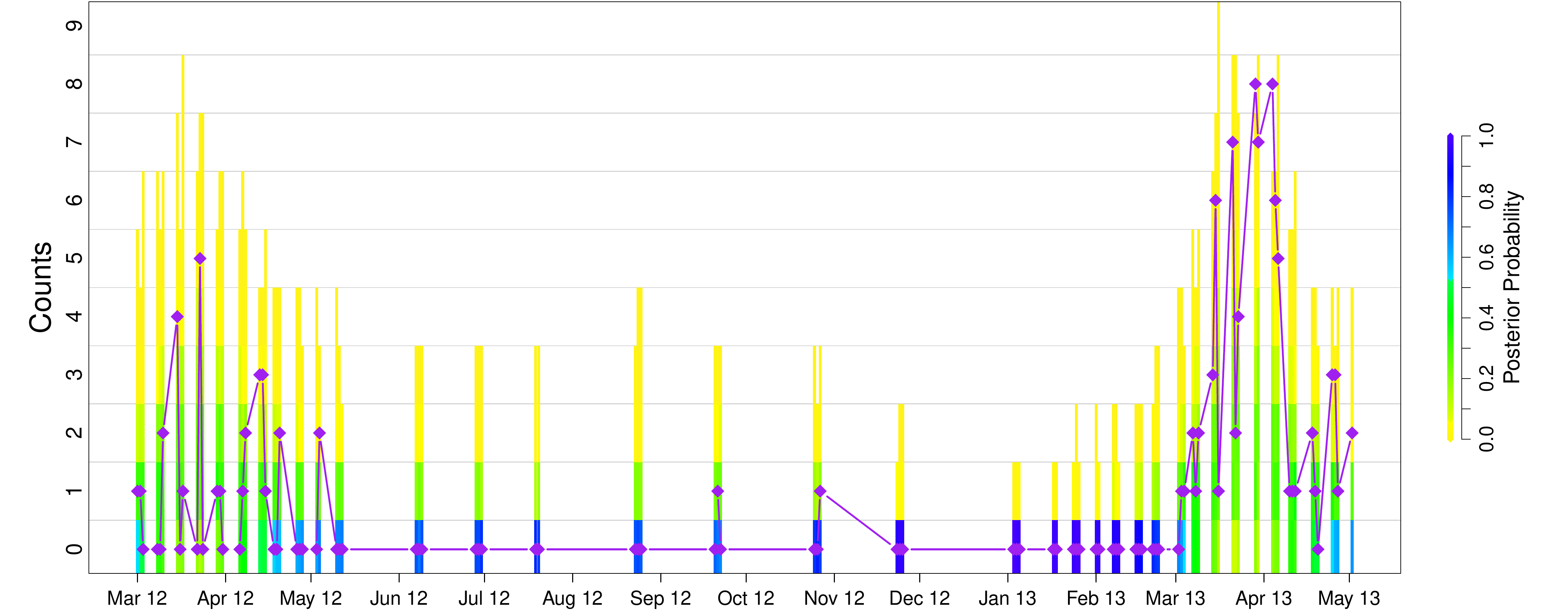}
    \includegraphics[width=1.1\linewidth]{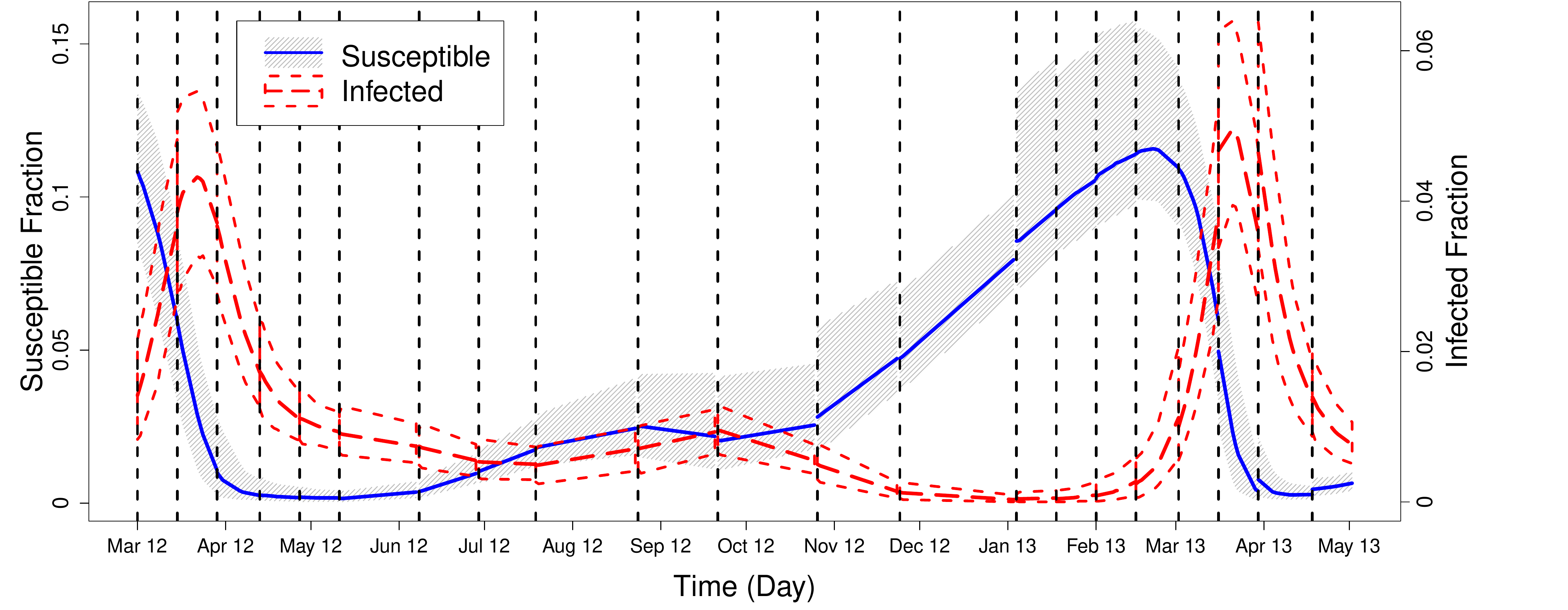}
\vspace{-20pt}
    \caption{Summary of prediction results for the second to last and
      last epidemic peaks in the Bangladesh data. We again run PMMH
      algorithms on training sets of the data, which are cut off at
      each of the dashed black lines in the bottom plot, and future cases
      are predicted until the next cut off. The top plot compares
      the posterior probability of the predicted counts to the test
      data (purple diamonds and line), and the bottom plot shows how
      the trajectory of the predicted hidden states changes over the
      course of the epidemic. See the caption of Figure
      \ref{predsimpred} for more details.  }
    \label{mathpred}
  \end{centering}
\end{figure}

For the data collected from Mathbaria, we begin prediction at multiple
points around the time of the two epidemic peaks that occur in 2012
and 2013. Figure \ref{cholbp} shows the full cholera data with
smoothed and standardized covariates. Figure \ref{mathpred} shows the
posterior predictive distribution of observed cholera cases (top row)
and \hidden states from the time-varying SIRS model (bottom
row). Parameters used to simulate the SIRS forward in time have been sampled
using the PMMH algorithm applied to the training data, with data being
cut off at different points during the 2012 and 2013 epidemic
peaks. From each of these cut offs, parameter values are then used to
simulate possible realizations of the test data. Predictions are run
until the next cut off point, with cut off points chosen based on the
length of the lag $\kappa$. Realistically, at time $t$ we have covariate information
to use for prediction only until time $t+\kappa$, where $\kappa$ is
the covariate lag. Since the smallest lag considered is 14 days, we
make only 14 day ahead predictions where possible to mimic a realistic prediction set up. 
Due to the sparse sampling between epidemic peaks (June 2012 to February 2013), we use
longer prediction intervals for these cut-offs than would be possible
in real time data analysis in order to evaluate our model predictions.

In the top row of Figure \ref{mathpred}, the coloring of the bars
again represents the distribution of predicted cases. Between the two
peaks of case counts (June 2012 to February 2013), the frequency of
predicted zero counts is very high, so we conclude that the model is
doing well with respect to predicting the lack of an epidemic. During
the epidemics, the distribution of the counts shifts its mass away
from zero. The plot in the bottom row of Figure \ref{mathpred} again
illustrates the periodic nature and interplay of the \hidden states of
the underlying compartmental model. When the fraction of infected
individuals quickly increases during an epidemic, the fraction of
susceptibles decreases. Afterwards, the fraction of infected
individuals drops to almost zero and the pool of susceptibles is
slowly replenished. When the fraction of infected individuals is low,
there is more uncertainty in the prediction for the fraction of
susceptibles (September 2012 to March 2013). The fraction of infected
individuals increases to a slightly higher epidemic peak 2013 (March
2013 to May 2013) than in 2012 (March 2012 to May 2012), as observed
in the test data for those years. The predicted fraction of infected
people in the population increases before an increase can be seen in
the case counts, which could allow for early warning of an epidemic.

We also use a quasi-Poisson regression model similar to the one used
by \cite{phase1} to predict the mean number of cholera cases (Appendix
E).  Although the quasi-Poisson model predicts reasonably well the
timing of epidemic peaks, it appears to overestimate the duration of
the outbreaks. The predicted means under both the quasi-Poisson and
SIRS models most likely underestimate the true mean of the observed
counts, with the quasi-Poisson model performing slightly better.
However, the SIRS predicted fraction of infected individuals --- a
hidden variable in the SIRS model --- provides a more detailed picture
of how cholera affects a population. By providing not only accurate
prediction of the time of epidemic peaks, but also the predicted
fraction of the population that is infected, the SIRS model
predictions could be used for efficient resource allocation to treat
infected individuals. See Appendix E for additional details.

\section{Discussion}

We use a Bayesian framework to fit a nonlinear dynamic model for
cholera transmission in Bangladesh which incorporates environmental
covariate effects. We demonstrate these techniques on simulated data
from a hidden SIRS model with a time-varying environmental force of
infection, and the results show that we are recovering well the true
parameter values. We also estimate the effect of two environmental
covariates on cholera case counts in Mathbaria, Bangladesh while
accounting for infectious disease dynamics, and we test the predictive
ability of our model. Overall, the prediction results look
promising. Based on data collected, the predicted \hidden states show
a noticeable increase in the fraction of infected individuals weeks
before the observed number of cholera cases increases, which could
allow for early notification of an epidemic and timely allocation of
resources. The predicted \hidden states show that the fraction of
infected individuals in the population decreases greatly between
epidemics, supporting the hypothesis that the environmental force of
infection triggers outbreaks. Estimates of $\beta I_t$ are low, but
not negligible, compared to estimates of $\alpha(t)$, suggesting that
most of the transmission is coming from environmental sources.

Computational efficiency is an important factor in determining the
usefulness of this approach in the field. We have written an R package
which implements the PMMH algorithm for our hidden SIRS model,
available at \texttt{https://github.com/vnminin/bayessir}. The
computationally expensive portions of the PMMH code are primarily
written in C++ to optimize performance, using Rcpp to integrate C++
and R \citep{Rcpp,Rcppbook}; however there is still room for
improvement. Running $400000$ iterations of the PMMH algorithm on the
six years of data from Mathbaria takes 2.5 days on a 4.3 GHz i7
processor. Since we can predict three weeks into the future using a 21
day covariate lag, we do not think timing is a big limitation for
using our model predictions in practice.

Plots of residuals over time, shown in Appendix C, show that we are
modeling well case counts between the epidemic peaks but not the
epidemic peaks themselves, either due to missing the timing of the
epidemic peak or the latent states not being modeled accurately. This
possible model misspecification might be fixed by including more
covariates, using different lags, or modifying the SIRS model. Also,
we assume a constant reporting rate, $\rho$, rather than using a
time-varying $\rho_t$ \citep{TSIR2000}. With better quality data we
might be able to allow for a reporting rate that varies over time; we
will try to address these model refinements in future analyses.

In the future, we will extend this analysis to allow for variable
selection over a large number of covariates. This will allow us to
include many covariates at many different lags and incorporate
information from all of the water bodies in a way that does not
involve averaging. In the current PMMH framework, choosing an optimal
proposal distribution to explore a much larger parameter space would
be difficult. We want to include a way of automatically selecting
covariates or shrinking irrelevant covariate effects to zero with
sparsity inducing priors. The particle Gibbs sampler, introduced by
\cite{pmcmclong}, would allow for such extensions. Approximate
Bayesian computation is also an option for further model development
\citep{ABCepi}.  In addition, the available data consist of
observations from multiple thanas during the same time period. Future
analyses will look into sharing information across space and time and
accounting for correlations between thanas. Another challenging future
direction involves exploring models which incorporate a feedback loop
from infected individuals back into the environment to capture the
effect of infected individuals excreting \textit{V. cholerae} into the
environment. To accomplish this, we could add a water compartment to
our SIRS model that quantifies the concentration of
\textit{V. cholerae} in the environment, similar to the model of
\cite{tien}. However, adding an additional latent state leads to
identifiability problems, even with fully observed data
\citep{eisen2013}, so such an extension will require rigorous testing
and fine tuning.

\section*{Acknowledgments}
AAK, MEH, and IML were supported by the NIH grants R01-AI039129,
U01-GM070749, and U54-GM111274. VNM was supported by the NIH grant R01-AI107034. JW was
supported by the NIH grant R01-AI029168. The authors gratefully
acknowledge collaborators at the ICDDR,B who collected and processed
the data. 


\clearpage
\section*{Supplementary Materials}

\renewcommand{\thefigure}{A-\arabic{figure}}
\setcounter{figure}{0}
\renewcommand{\thetable}{A-\arabic{table}}
\setcounter{table}{0}
\section*{Appendix A: Simulation Details}

\subsection*{PMMH pseudocode}

The following exposition of the algorithm follows closely the
pseudocode of \cite{pmcmclong} and \cite{wilk2011}.

Step 1: initialization, for iteration $j=0$,
\begin{enumerate}[(a)]
\item Set $\bmtheta(0)$ arbitrarily
\item Run the following SMC algorithm to get
  $\hat{p}(\bmy|\bmtheta(0))$, an estimate of the marginal
  likelihood, and to produce a sample $\bmX_{\bmt_{0:n}}(0) \sim
  \hat{p}(\cdot|\bmy,\bmtheta(0))$.

  Let the superscript $k \in \{1,\ldots,K\}$ denote the particle
  index, where $K$ is the total number of particles, and the subscript
  $t_i \in \{t_0,\ldots,t_n\}$ denote the time; thus, $\bmX_{t_i}^k$
  denotes the $k$th particle at time $t_i$, and
  $\bmX_{\bmt_{0:i}}^k=\left(\bmX_{t_0}^k,\ldots,\bmX_{t_i}^k\right)$. At
  time $t_i=t_0$, sample $\bmX_{t_0}^k=(S_{t_0}^k,I_{t_0}^k)$ for $k =
  1, \ldots, K$ from the initial density of the hidden Markov state
  process. Specifically, sample $S_{t_0}^k \sim
  \mbox{Poisson}(\phi_{S})$ and $I_{t_0}^k \sim
  \mbox{Poisson}(\phi_{I})$. Compute the $k$ weights $w(\bmX_{t_0}^k)
  := \mbox{Pr}(y_{t_0}|\bmX_{t_0}^k,\bmtheta(0))={I_{t_0}^k \choose
    y_{t_0}} \rho(0)^{y_{t_0}}(1-\rho(0))^{I_{t_0}^k-y_{t_0}}$, and
  set $W(\bmX_{t_0}^k)=w(\bmX_{t_0}^k)/\sum_{k'=1}^K
  w(\bmX_{t_0}^{k'})$.

  For $i=1, \ldots, n$, resample $\bar{\bmX}_{t_{i-1}}^k$ from
  $\bmX_{t_{i-1}}^k$ with weights $W(\bmX_{t_{i-1}}^k)$. Sample $K$
  particles $\bmX_{t_{i}}^k$ from $p(\cdot|\bar{\bmX}_{t_{i-1}}^k)$
  (i.e. propagate resampled particles forward one time point). Assign
  weights $w(\bmX_{t_i}^k):=
  \mbox{Pr}(y_{t_i}|\bmX_{t_i}^k,\bmtheta(0))$ and compute normalized
  weights $W(\bmX_{t_i}^k)=w(\bmX_{t_i}^k)/\sum_{k'=1}^K
  w(\bmX_{t_i}^{k'})$. Set $\bmX_{\bmt_{0:i}}^k =
  (\bar{\bmX}_{\bmt_{0:i-1}}^k,\bmX_{t_i}^k)$.

  It follows that
  \[
  \hat{p}(y_{t_i}|\bmy_{\bmt_{0:i-1}},\bmtheta(0))=\frac{1}{K}\sum_{k=1}^Kw(\bmX_{t_i}^k)
  \]
  is an approximation to the likelihood $p(y_{t_i}|\bmy_{\bmt_{0:i-1}},\bmtheta(0))$,
  and therefore an approximation to the total likelihood is
  \[
  \hat{p}(\bmy|\bmtheta(0))=\hat{p}(y_{t_0}|\bmtheta(0))
  \prod_{i=1}^n\hat{p}(y_{t_i}|\bmy_{\bmt_{0:i-1}},\bmtheta(0)).
  \]

  Thus we have a simple, sequential, likelihood-free algorithm which
  generates an unbiased estimate of the marginal likelihood,
  $p(\bmy|\bmtheta(0))$. A $\bmX_{\bmt_{0:n}}(0)$ trajectory is
  sampled from the $K$ trajectories ($\bmX_{\bmt_{0:n}}^k$, for
  $k=1,\ldots,K$) based on the final set of particle weights,
  $W(\bmX_{t_n}^k)$.
\end{enumerate}

Step 2: for iteration $j \geq 1$,
\begin{enumerate}[(a)]
\item Sample $\bmtheta^* \sim q\{ \cdot | \bmtheta(j-1)  \}$ 
\item Run an SMC algorithm, as in step 1(b) with $\bmtheta^*$ instead
  of $\bmtheta(0)$, to get $\hat{p}(\bmy|\bmtheta^*)$ and
  $\bmX^*_{\bmt_{0:n}} \sim \hat{p}(\cdot|\bmy,\bmtheta^*)$
\item With probability 
\[
\mbox{min} \left\{ 1,
\frac{\hat{p}(\bmy|\bmtheta^*)}{\hat{p}(\bmy|\bmtheta(j-1))}
\frac{\Prob(\bmtheta^*)}{\Prob\{ \bmtheta(j-1) \}} \frac{q\{\bmtheta(j-1)
  |\bmtheta^*\}}{q\{\bmtheta^*|\bmtheta(j-1) \}} \right\}
\]
set $\bmtheta(j)=\bmtheta^*$, $\bmX_{\bmt_{0:n}}(j)= \bmX^*_{\bmt_{0:n}}$, and
$\hat{p}(\bmy|\bmtheta(j))=\hat{p}(\bmy|\bmtheta^*)$, otherwise
set $\bmtheta(j)=\bmtheta(j-1)$, $\bmX_{\bmt_{0:n}}(j)= \bmX_{\bmt_{0:n}}(j-1)$, and
$\hat{p}(\bmy|\bmtheta(j))=\hat{p}(\bmy|\bmtheta(j-1))$.
\end{enumerate}

\subsection*{Simulating homogeneous SIRS}

Gillespie's direct method \citep{gill1977} simulates the time to the
next event and then determines which event happens at that time. The
first reaction method \citep{gill1976} calculates the time to the next
reaction for each of the possible events, and the minimum time to next
reaction determines the next step of the chain.

Using the direct method, we can think of our continuous-time
Markov chain (CTMC) as a chemical system
with three different reactions. These reactions and their rate
functions are given by the infinitesimal rates
\begin{eqnarray*}
  \lambda_{(S,I,R),(S',I',R')} = \left\{ \begin{array}{rl}
      (\beta I + \alpha) S &\mbox{ if $S'=S-1$, $I' = I+1$, $R'=R$}, \\
      \gamma I &\mbox{ if $S'=S$, $I' = I-1$, $R'=R+1$}, \\
      \mu R &\mbox{ if $S'=S+1$, $I' = I$, $R'=R-1$}, \\
      0 &\mbox{ otherwise.}
    \end{array} \right.
\end{eqnarray*}
Thus the three reactions have the rate functions
$h_1(\bmX_t)=(\beta I_t+\alpha) S_t$, $h_2(\bmX_t) = \gamma I_t$,
and $h_3(\bmX_t) = \mu R_t$, corresponding to the infinitesimal
rates of the CTMC. Then the time to the next reaction, $\tau$,
has an exponential distribution with rate $\lambda =
h_1(\bmX_t)+h_2(\bmX_t)+h_3(\bmX_t)$, and the $k$th reaction occurs
with probability $h_k(\bmX_t)/ \lambda$, for $k = \{1,2,3\}$.

The first reaction method instead simulates the time $\tau_k$ that the
$k$th reaction happens for $k = \{1,2,3\}$, given no other reactions
happen in that time. Then the time to the next reaction
$\tau=\mbox{min}_k(\tau_k)$, and the reaction with the reaction time equal to
$\tau$ is the event that happens. 

Both the direct method and the first reaction method work only for
homogeneous Markov chains.  If we want to assume that the additional
force of infection, $\alpha$, varies over time, the associated Markov
chain is inhomogeneous and we must account for the fact that the
transition rate could change before the next reaction occurs.

\subsection*{Simulating inhomogeneous SIRS}

\citet{nrm} introduce the next reaction method, an efficient exact
algorithm to simulate stochastic chemical systems. They extend this
next reaction method to include time-dependent rates and non-Markov
processes. \citet{modgill} deviates from these methods a bit, using
Poisson processes to represent the reaction times, with time to next
reaction given by integrated rate functions. This leads to a more
efficient modified next reaction method which they extend to systems
with more complicated reaction dynamics.

Using the methods described by \cite{nrm} and \cite{modgill}, to
incorporate a time-varying force of infection into the SIRS model we
must integrate over the rate function $h_1(\bmX_t,s)=\left(\beta
  I_t+\alpha(s) \right)S_t$. Thus, to find the time $\tau_1$ that the
first reaction happens, given no other reactions happen in that time,
we generate $u \sim \mbox{Uniform}(0,1)$ and solve

\begin{eqnarray*}
\int_t^{\tau_1} h_1(\bmX_t,s) ds = \mbox{ln}(1/u)
\end{eqnarray*}

for $\tau_1$. Since the other two reactions have no time-varying
parameters, we can solve for $\tau_2$ and $\tau_3$, the reaction times
of the second and third reactions, using the methods of the previous
section. Then we can continue, using the first reaction method to
simulate the process.
 
\begin{figure}[!h]
  \begin{centering}
    \includegraphics[width=1\linewidth]{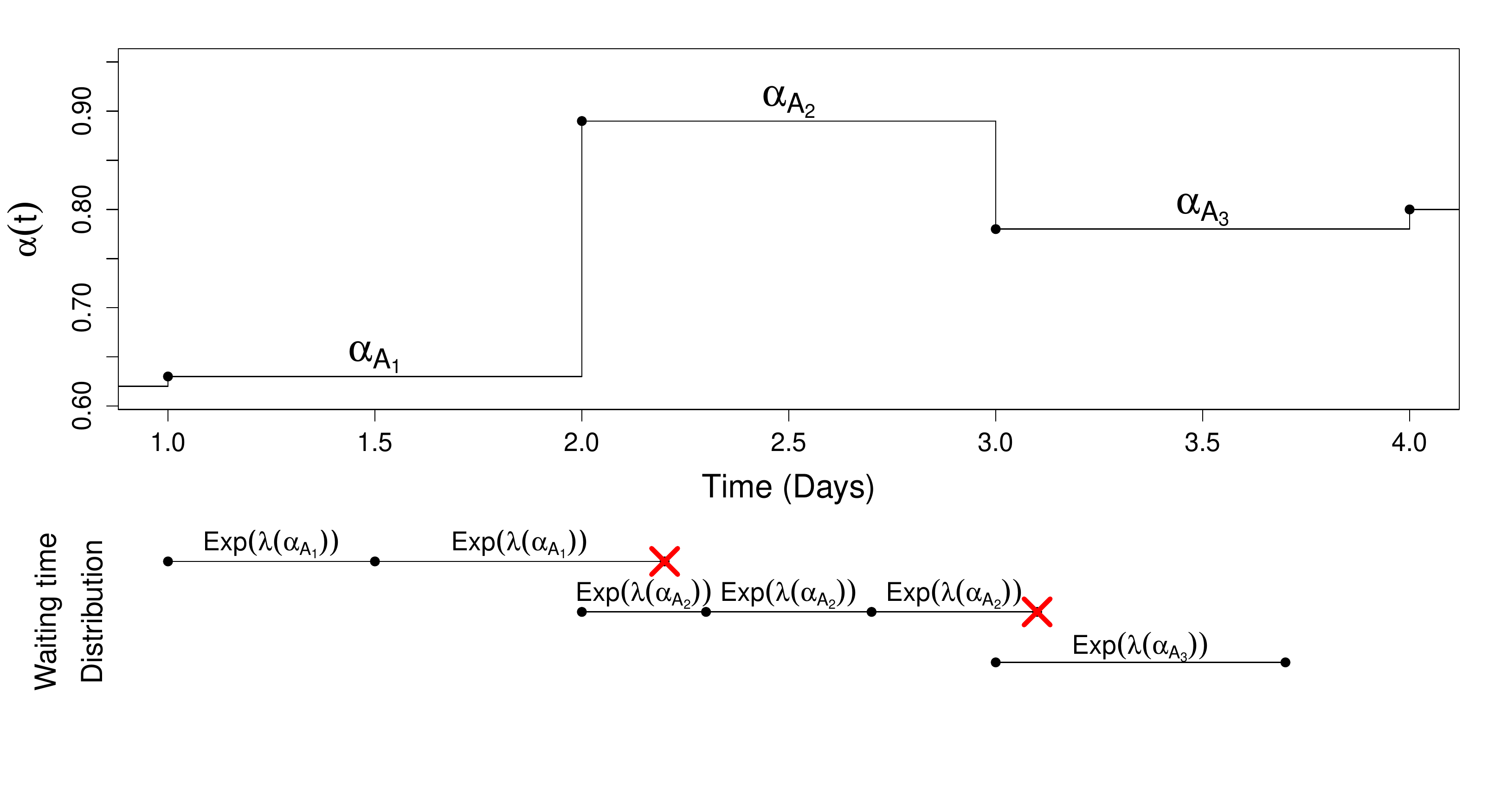}
    \caption{\small{Depiction of the modified Gillespie algorithm. We
        assume the environmental force of infection, $\alpha(t)$, is a
        step function which changes daily. Daily time intervals are
        denoted by $A_i := [i,i+1)$ for $i \in \{t_0,t_0+1,\ldots,t_n-1\}$,
        so $\alpha(t)=\alpha_{A_i}$ for $t \in A_i$. Starting at time
        $t=1$, the waiting time to the next event, $\tau$, has an
        exponential distribution with rate
        $\lambda(\alpha_{A_1})=h_1(\bmX_t,\alpha_{A_1})+h_2(\bmX_t)+h_3(\bmX_t)$. In
        the depiction, $\tau=0.5$. The simulated waiting time plus the
        current time, $t^*=t+\tau$, remains in the interval $A_1$, so
        we use $t^*$ as the next time in our CTMC and propagate
        $\bmX_t$ forward at that time using Gillespie's direct
        method. Since we are still in the interval $A_1$, we again
        simulate the time to the next event as an exponential random
        variable with rate
        $\lambda(\alpha_{A_1})=h_1(\bmX_{t^*},\alpha_{A_1})+h_2(\bmX_{t^*})+h_3(\bmX_{t^*})$.
        In this iteration, the waiting time plus the current time,
        $t^*+\tau$, exceeds the boundary of the interval $A_1$, so we
        discard this simulated waiting time $\tau$. Using the
        memoryless property of exponentials, we restart our simulation
        from the beginning of the interval $A_2$ using the new
        $\alpha(t)$ value, $\alpha_{A_2}$. We continue in this manner
        until we have simulated the Markov process $\bmX_t$ up to time
        $t_n$.}}
    \label{modgillPlot}
  \end{centering}
\end{figure}

We simplify this approach by assuming that the time-varying force of
infection, $\alpha(t)$, remains constant each day. We define daily
time intervals $A_i := [i,i+1)$ for $i \in
\{t_0,t_0+1,\ldots,t_n-1\}$, and $\alpha(t)=\alpha_{A_i}$ for $t \in
A_i$. Then we can take advantage of the memoryless property of
exponentials and propagate the chain forward in daily
increments. Thus, we use the direct method, but when the time to next
event exceeds the right end point of the current interval $A_i$, we
restart CTMC simulation from the beginning of the interval $A_{i+1}$
using $\alpha_{A_{i+1}}$ in the waiting time distribution rate
$\lambda(\alpha_A)=h_1(\bmX_t,\alpha_A)+h_2(\bmX_t)+h_3(\bmX_t)$, so
$\tau \sim \mbox{Exp}(\lambda(\alpha_A))$. This modified Gillespie
algorithm is depicted and detailed in Figure \ref{modgillPlot}.

\subsection*{Selecting Tau}

\begin{figure}[h]
  \begin{centering}
    \includegraphics[width=1\linewidth]{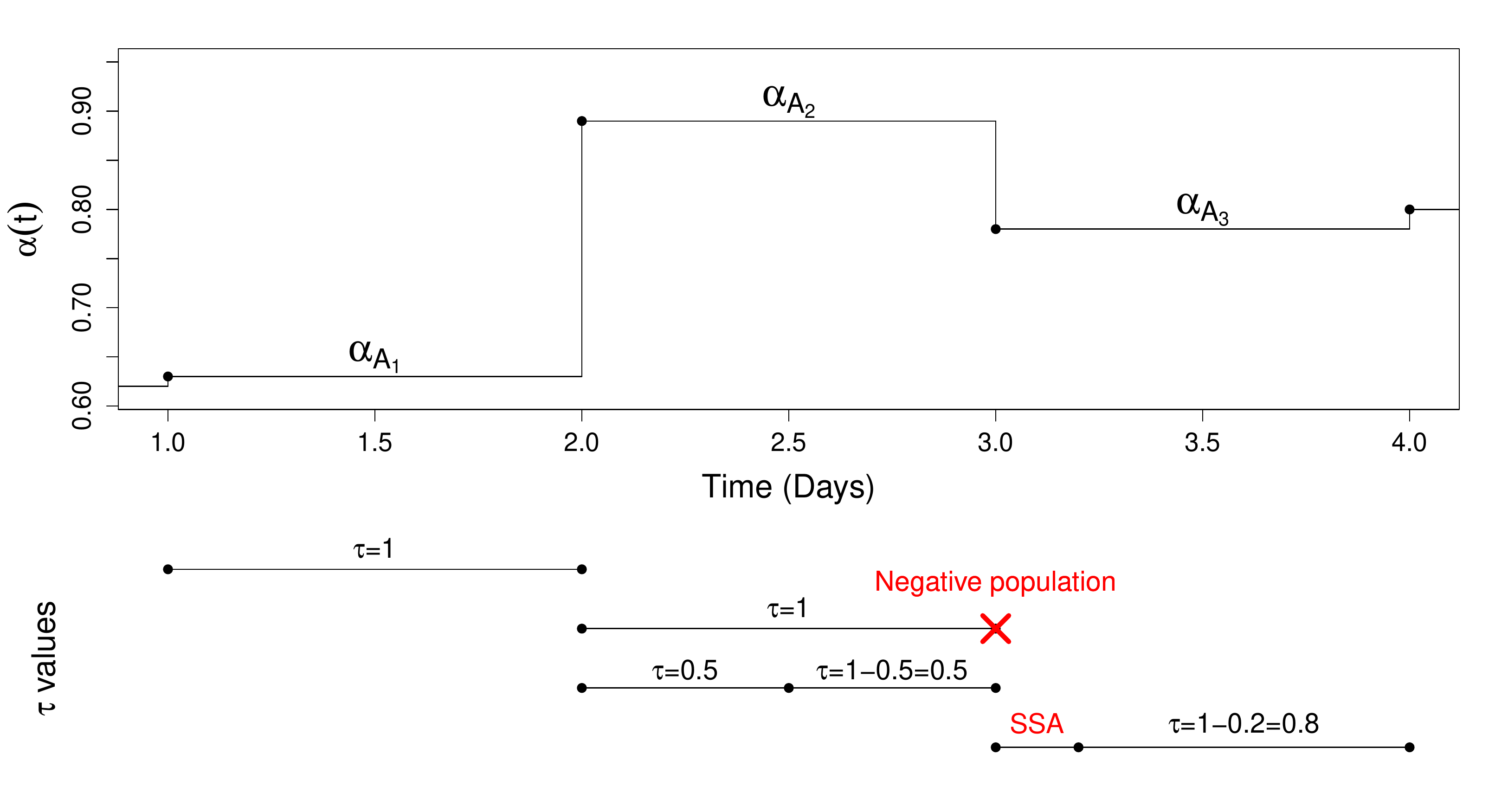}
    \caption{Depiction of the modified tau-leaping algorithm. We
      assume the environmental force of infection, $\alpha(t)$, is a
      step function which changes daily. Daily time intervals are
      denoted by $A_i := [i,i+1)$ for $i \in \{t_0,t_0+1,\ldots,t_n-1\}$, so
      $\alpha(t)=\alpha_{A_i}$ for $t \in A_i$. As a default, we use
      $\tau=1$ day. Starting at time $t=1$, we simulate the changes in
      compartment populations over the interval $t \in [1,2)$. At time
      $t=2$, we again use $\tau=1$ day to simulate the changes over
      the interval $t \in [2,3)$. This value of $\tau$ produces a
      negative population so we reject that simulation and try again
      with a smaller $\tau$ (reduced by a factor of $1/2$). The next
      value of $\tau$ is then calculated based on how long the current
      daily time-varying force of infection remains constant, so
      $\tau=0.5$. At time $t=3$, the population of a compartment is
      lower than some prespecified critical size, so a single step
      algorithm (SSA), in our case the Gillespie algorithm, is used
      until the population gets above that critical size. Once the
      compartment populations are all above the critical size again,
      at time $t=3.2$, the subsequent value of $\tau$ is again picked
      based on how long the current daily time-varying force of
      infection remains constant, so $\tau=0.8$.}
    \label{picktau}
  \end{centering}
\end{figure}

Unchecked, tau-leaping can lead to negative population sizes in a
compartment if the compartment has a low number of individuals. To
avoid this, we use a simplified version of the modified tau-leaping
algorithm presented by \cite{gillmodTL}. If the population of a
compartment is lower than some prespecified critical size, a single
step algorithm (like the Gillespie algorithm) is used until the
population gets above that critical size. If the size of the
compartment is not critically low but the current value of $\tau$
still produces a negative population, we reject that simulation and
try again with a smaller $\tau$ (reduced by a factor of $1/2$).  The
subsequent value of $\tau$ is picked based on how long the current
daily time-varying force of infection remains constant. We choose a
value of $\tau$ that simulates what happens during the remainder of
the day, until the value of the transition rate changes. This
modified tau-leaping algorithm is depicted and detailed in Figure
\ref{picktau}.

\begin{figure}[h]
  \begin{centering}
    \includegraphics[width=.7\linewidth]{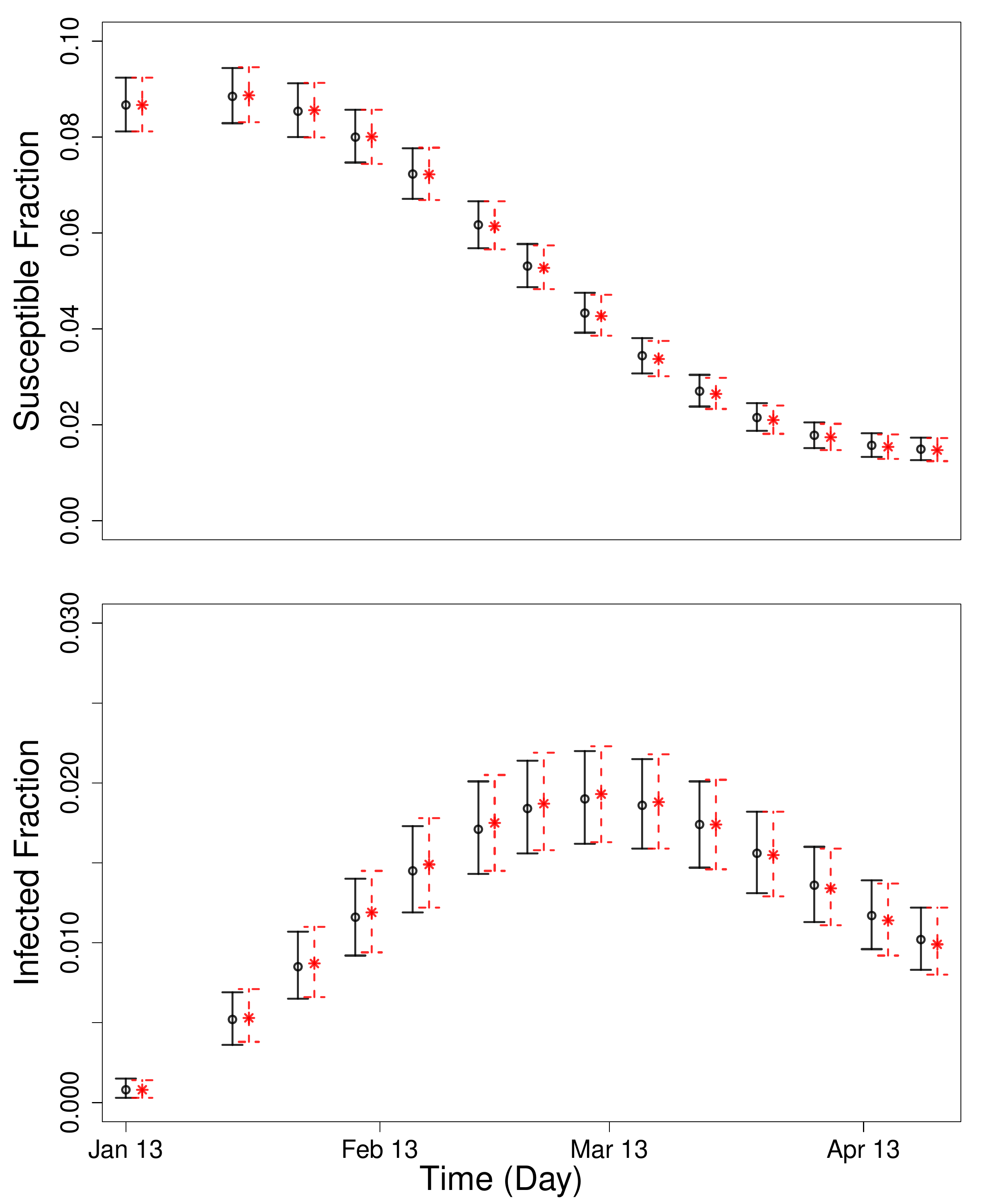}
    \caption{Plots comparing the median and 95\% intervals at
      different points during an epidemic, simulated using both the
      modified Gillespie algorithm and the modified tau-leaping
      algorithm with $\tau=1$ day. The medians and 95\% intervals for
      5000 simulations using the Gillespie algorithm are given by the
      open circle and solid error bars. The medians and
      95\% intervals for 5000 simulations using the modified
      tau-leaping algorithm are given by the asterisk and dashed error
      bars. }
    \label{TLcomp}
  \end{centering}
\end{figure}

For our simulations, we have chosen $\tau=1$ day; we perform a
simulation study to see if this value for $\tau$ is reasonable. 
Using the posterior estimates of the parameters, we simulate data forward in time 5000
times using both the modified Gillespie algorithm and the modified
tau-leaping algorithm. We simulate data over the entire epidemic curve
to see how the comparison changes for varying values of $\alpha(t)$. 
Figure \ref{TLcomp} shows estimates of the median and 95\% intervals
for the simulated values. The Monte Carlo standard error is very small for all estimates.  For
the numbers of susceptible individuals, the estimates under Gillespie
and tau-leaping are almost identical over the entire epidemic. For the
numbers of infected, the values are very close except at the epidemic
peaks. However, the differences are very small. We conclude that for
our application $\tau=1$ day is a good compromise between
computational efficiency and accuracy.

\subsection*{Binomial tau-leaping}
Another solution to the negative population size problem is to use
Binomial tau-leaping \citep{chatterjeeBinomial,tianBinomial}, which
further approximates $k_j$ as a binomial random variable with mean
$h_j(\bmX_t)\tau$ and upper limit chosen such that $k_j$ cannot be
large enough to simulate a negative population. We opt instead to
use the simplified version of the modified tau-leaping algorithm.

\clearpage

\renewcommand{\thefigure}{B-\arabic{figure}}
\setcounter{figure}{0}
\renewcommand{\thetable}{B-\arabic{table}}
\setcounter{table}{0}
\section*{Appendix B: MCMC diagnostics}
\begin{figure}[h!]
  \begin{centering}
 \includegraphics[width=1\linewidth]{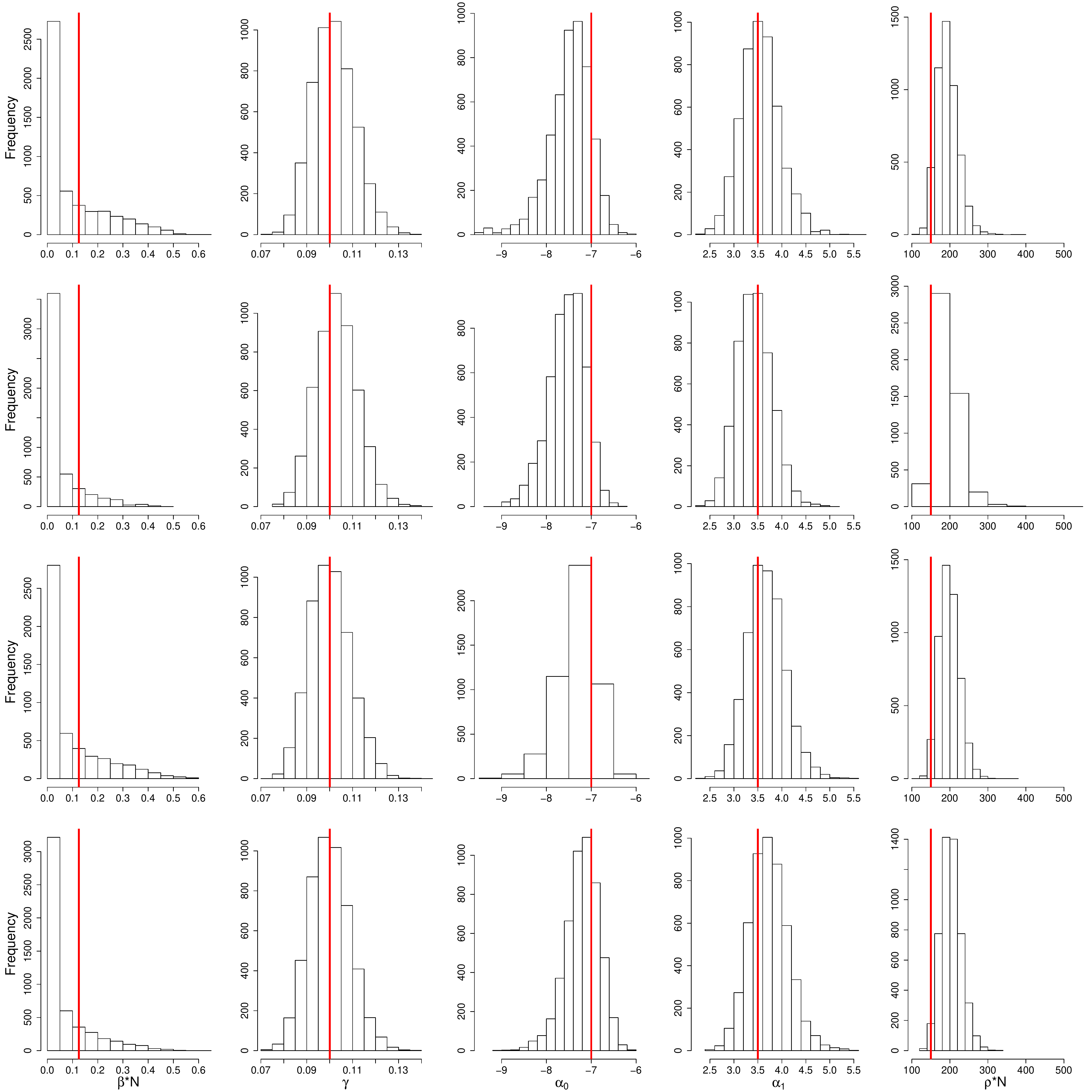}
 \caption{Posterior distributions for the parameters of the SIRS
   model, based on simulated data. From top to bottom, the rows have
   $\phi_S/N$ and $\phi_I/N$ above the true values (0.31 and 0.003),
   at the true values (0.21 and 0.0015), below the true values (0.11
   and 0.00075), and further below (0.055 and 0.000375). The true
   values of the parameters are denoted by the red lines.}
    \label{posthist}
  \end{centering}
\end{figure}

\begin{figure}[h!]
  \begin{centering}
  \includegraphics[width=1\linewidth]{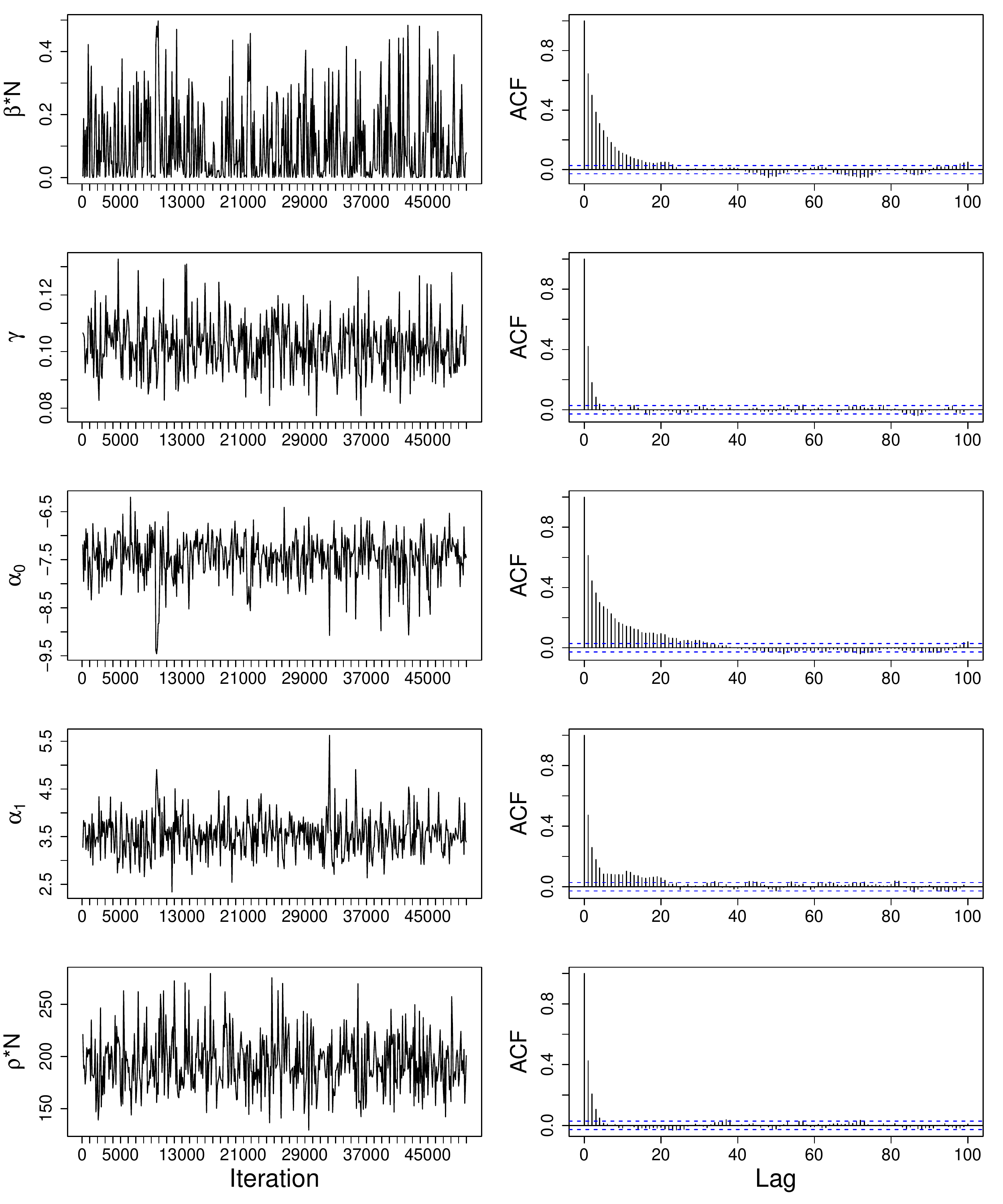}
  \caption{Summary plots of the PMMH algorithm output (final run of
    $50000$ iterations) for the parameters of the SIRS model, based on
    simulated data. ACF plots are thinned to 5000 iterations and trace
    plots are thinned to display only 500 iterations.}
  \label{simresults}
  \vspace{-5pt}
  \end{centering}
\end{figure}

\begin{figure}[h!]
  \begin{centering}
    \includegraphics[width=1\linewidth]{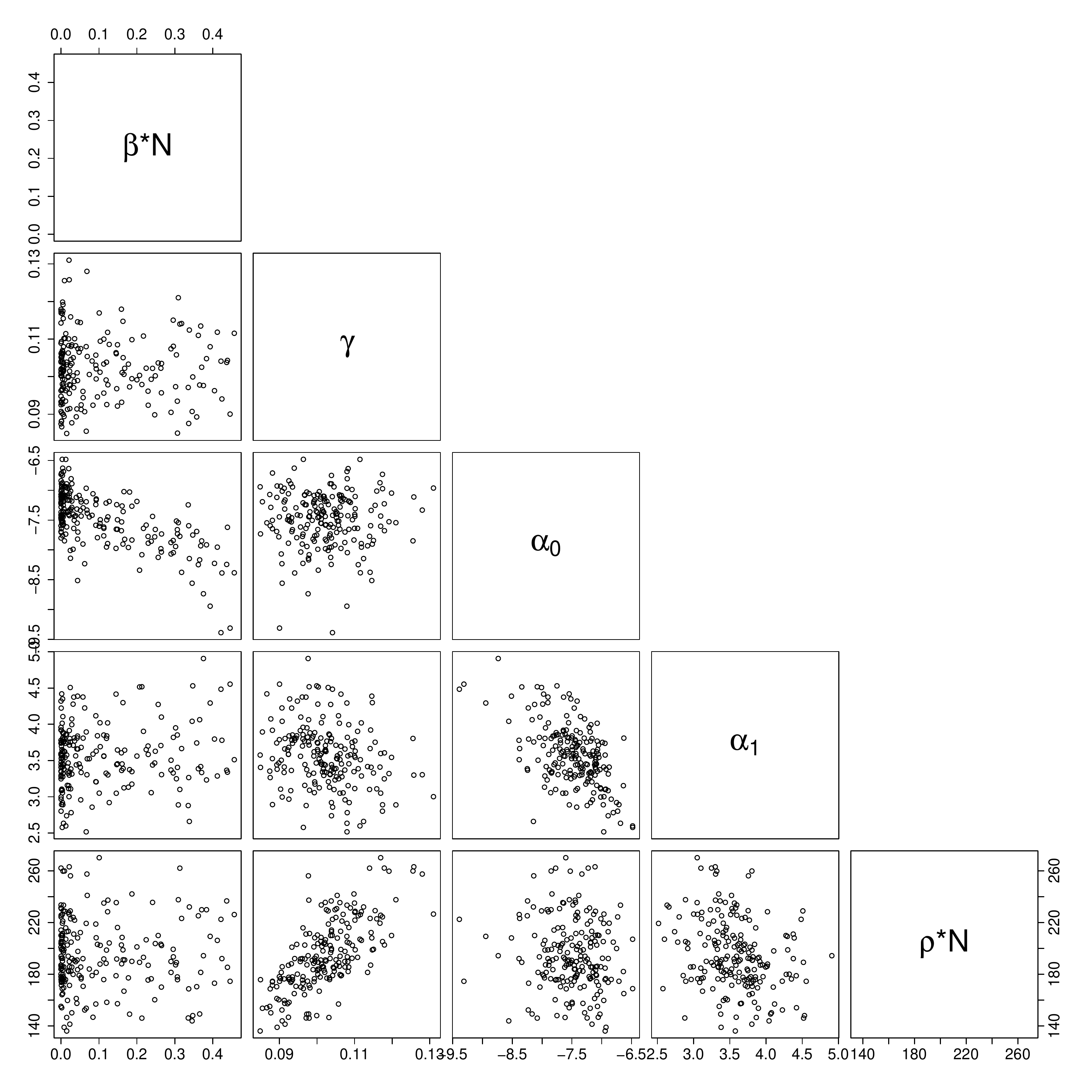}
    \caption{Bivariate scatterplots of parameters of the SIRS model,
      based on simulated data. Scatterplots are thinned to display
      only 200 samples, so only every 250th sample from the posterior
      distribution is plotted.}
  \label{simbivar}
  \vspace{-5pt}
  \end{centering}
\end{figure}

\begin{figure}[h]
  \begin{centering}
    \includegraphics[width=.95\linewidth]{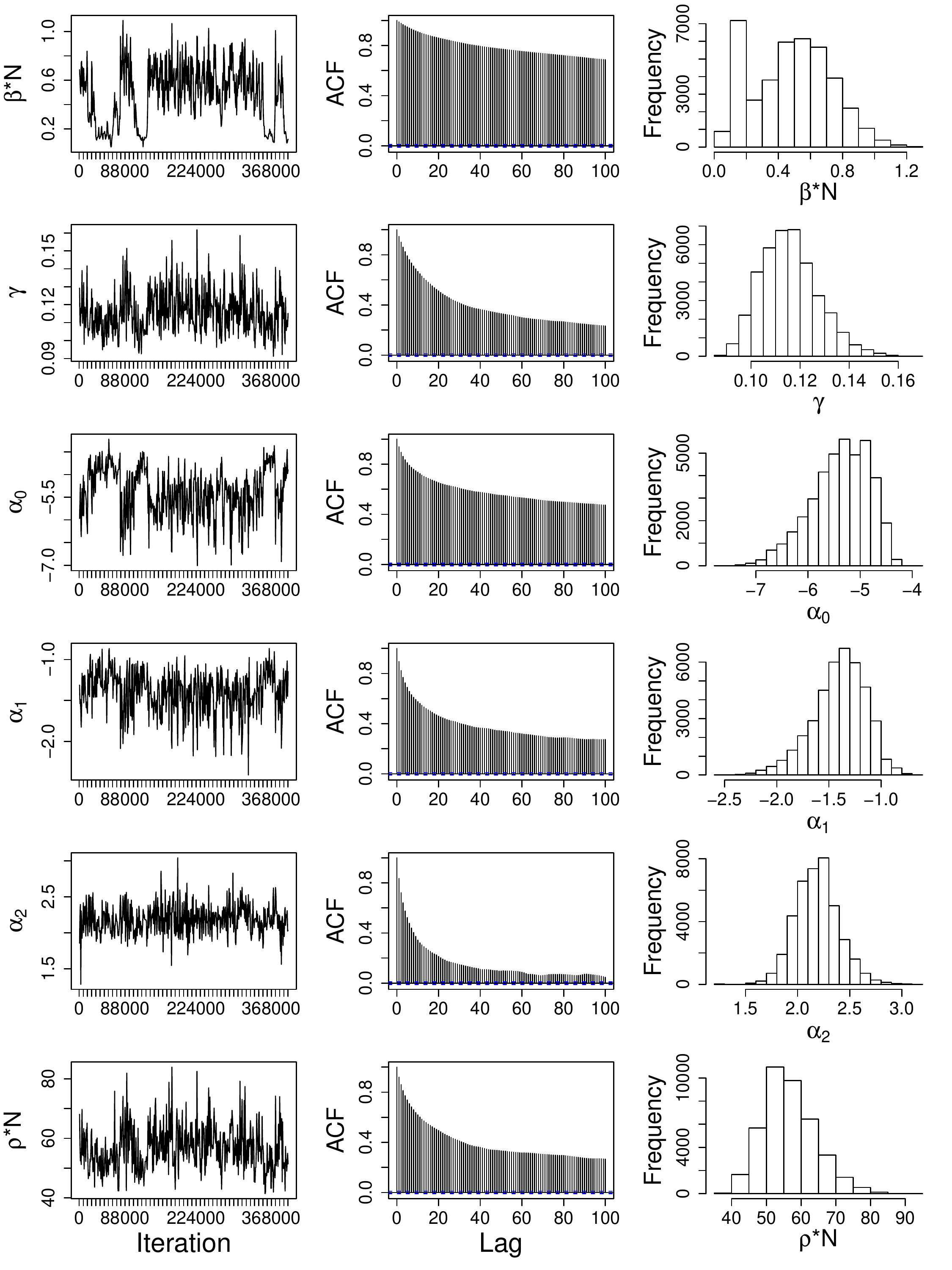}
    \caption{\small{Summary plots of the PMMH algorithm output (final
        run of $400000$ iterations) for the parameters of the SIRS
        model, based on data from Mathbaria, Bangladesh. ACF plots and
        histograms are thinned to 40000 iterations and trace plots are
        thinned to display only 500 iterations. }}
    \label{PMMHrealfinal}
  \end{centering}
\end{figure}

\begin{figure}[h!]
  \begin{centering}
    \includegraphics[width=.95\linewidth]{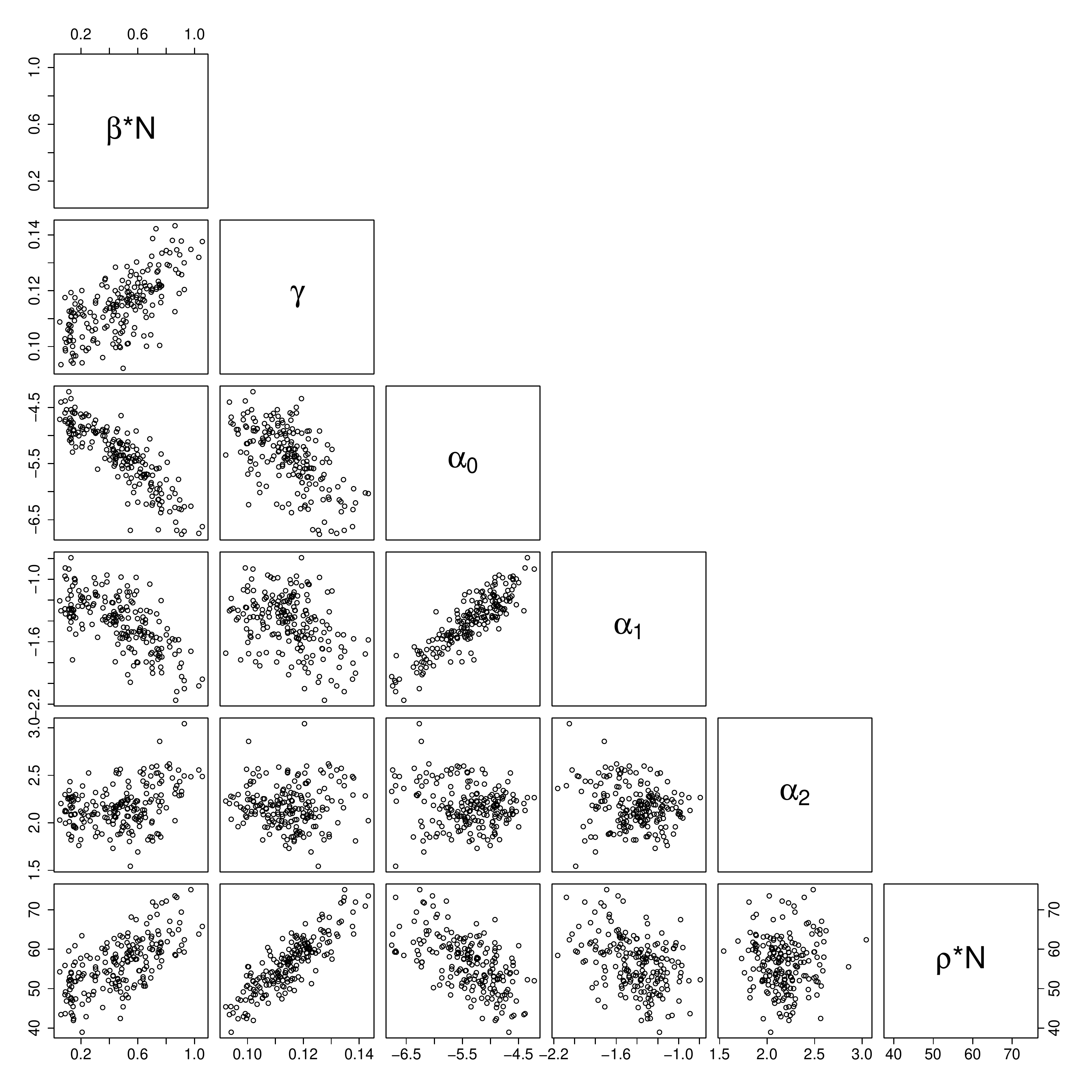}
    \caption{Bivariate scatterplots of parameters of the SIRS model
      estimated using data from Mathbaria. Scatterplots are thinned to
      display only 200 samples, so only every 2000th sample from the
      posterior distribution is plotted.}
    \label{PMMHrealfinalbivar}
  \end{centering}
\end{figure}

Using simulated data, we compare results from models with different
assumptions on the values of $\phi_S$ and $\phi_I$; marginal posterior
distributions for the parameters of the SIRS model from the final runs
of PMMH algorithms are in Figure \ref{posthist}. The posterior
distributions are similar, regardless of assumptions about $\phi_S/N$
and $\phi_I/N$. Trace plots and autocorrelation plots for the
parameters of the SIRS model assuming $\phi_S$ and $\phi_I$ are set at
the true values are in Figure \ref{simresults}, and Figure
\ref{simbivar} shows bivariate scatterplots of the parameters. Summary
plots of the PMMH algorithm output for the parameters of the SIRS
model with data from Mathbaria, Bangladesh are given in Figure
\ref{PMMHrealfinal}, and Figure \ref{PMMHrealfinalbivar} shows
bivariate scatterplots of the parameters. Effective sample sizes range
from 593 to 2038 for the parameters of the SIRS model with a
time-varying environmental force of infection and from 77 to 1545 for
the analysis of the data from Mathbaria. To test convergence, we
varied the initial values for the parameters of the PMMH
algorithm. Some of the initial values are shown in Table \ref{SVs} and
the parameter estimates from the chains that started at these initial
values are given in the top third of Table \ref{RDests}. Credible
intervals for $\beta \times N$ vary slightly for different initial
values; this is likely due to a heavy tail in the posterior
distribution that is not yet explored in the run initializing from the
second set of starting value and is most often explored in the run
initializing at the third set of starting values. If we obtained
larger samples from the posterior, the credible intervals would be
more similar.

\begin{table}[B]
  \centering
  \begin{tabular}{rlll}
    Coefficient  & Starting value set 1 & Starting value set 2 & Starting value set 3\\
    \hline
    $\beta \times N$ & 0.6 & 0.8 & 0.06 \\ 
    $\gamma$    &   0.11 & 0.1 & 0.12 \\ 
    $\alpha_0$ &   $-7.11$ & $-8$ & $-3$ \\ 
    $\alpha_1$   &   0 & 0 & 1 \\ 
    $\alpha_2$ &   0 & 0 & $-1$ \\ 
    $\rho \times N$ &   60 & 6 & 100 \\ 
    \hline
  \end{tabular}
  \caption{Initial values used for separate runs of the PMMH algorithm on the data from Mathbaria. We assume $N=10000$.}
  \label{SVs}
\end{table}

\begin{table}[h]
  \begin{centering}
    \begin{tabular}{rll@{,\ }rll@{,\ }rll@{,\ }rll@{,\ }r}
      & \multicolumn{3}{c}{Starting value set 1} & \multicolumn{3}{c}{Starting value set 2} & \multicolumn{3}{c}{Starting value set 3} &  \\\cline{2-10}
      Coefficient & Estimate & \multicolumn{2}{c}{95\% CIs} & Estimate & \multicolumn{2}{c}{95\% CIs} & Estimate & \multicolumn{2}{c}{95\% CIs} 
      \tabularnewline
      \hline 
      $\beta \times N$ & 0.49 & (0.1 & 0.95) & 0.53 & (0.22 & 0.97) & 0.38 & (0.02 & 0.91) \\ 
      $\gamma$ &       0.11 & (0.1 & 0.14) & 0.12 & (0.1 & 0.14) & 0.11 & (0.09 & 0.14) \\ 
      $(\beta \times N)/\gamma$ &     4.35 & (0.99 & 7.15) & 4.66 & (2.05 & 7.38) & 3.48 & (0.13 & 6.97) \\ 
      $\alpha_0$ &     -5.32 & (-6.63 & -4.51) & -5.43 & (-6.67 & -4.71) & -5.12 & (-6.46 & -4.45) \\  
      $\alpha_1$ &     -1.37 & (-1.98 & -0.98) & -1.41 & (-2.02 & -1.04) & -1.32 & (-1.89 & -0.95) \\  
      $\alpha_2$ &    2.18 & (1.8 & 2.62) & 2.19 & (1.8 & 2.67) & 2.17 & (1.8 & 2.57) \\ 
      $\rho \times N$ &      55.8 & (43.4 & 73.5) & 57 & (45.3 & 73.7) & 54.6 & (43.9 & 71.1) \\  
      \hline
      \hline
      & \multicolumn{3}{c}{N=10000, $1/\mu=2$ years} & \multicolumn{3}{c}{N=5000, $1/\mu=3$ years} & \multicolumn{3}{c}{N=50000, $1/\mu=3$ years}  \\
      & \multicolumn{3}{c}{$\phi_S/N=0.2$, $\phi_I/N=0.02$} & \multicolumn{3}{c}{$\phi_S/N=0.2$, $\phi_I/N=0.02$} & \multicolumn{3}{c}{$\phi_S/N=0.2$, $\phi_I/N=0.02$}  \\\cline{2-10}
      Coefficient & Estimate & \multicolumn{2}{c}{95\% CIs} & Estimate & \multicolumn{2}{c}{95\% CIs} & Estimate & \multicolumn{2}{c}{95\% CIs}  
      \tabularnewline
      \hline
      $\beta \times N$&0.55 & (0.36 & 0.71) & 0.74 & (0.35 & 1.04) & 0.8 & (0.23 & 1.04) \\ 
      $\gamma$ &       0.12 & (0.1 & 0.14) & 0.12 & (0.1 & 0.15) & 0.13 & (0.11 & 0.15) \\ 
      $(\beta \times N)/\gamma$ &       4.55 & (3.21 & 5.46) & 5.94 & (3.28 & 8.07) & 6.1 & (2.01 & 7.66) \\ 
      $\alpha_0$  &      -6.34 & (-7.33 & -5.29) & -6.12 & (-7.8 & -4.91) & -6.21 & (-7.18 & -4.95) \\ 
      $\alpha_1$ &      -1.83 & (-2.38 & -1.29) & -1.71 & (-2.35 & -1.16) & -1.76 & (-2.34 & -1.19) \\ 
      $\alpha_2$ &     2.3 & (1.8 & 2.85) & 2.14 & (1.34 & 2.74) & 2.38 & (1.87 & 2.97) \\ 
      $\rho \times N$ &     43.8 & (35.6 & 53.9) & 61.5 & (47.7 & 78.7) & 65.8 & (51.5 & 81.7) \\   
      \hline
      \hline
      & \multicolumn{3}{c}{N=10000, $1/\mu=3$ years} & \multicolumn{3}{c}{N=10000, $1/\mu=3$ years} & \multicolumn{3}{c}{N=10000, $1/\mu=3$ years}  \\
      & \multicolumn{3}{c}{$\phi_S/N=0.4$, $\phi_I/N=0.04$} & \multicolumn{3}{c}{$\phi_S/N=0.1$, $\phi_I/N=0.01$} & \multicolumn{3}{c}{$\phi_S/N=0.4$, $\phi_I/N=0.01$}  \\\cline{2-10}
      Coefficient & Estimate & \multicolumn{2}{c}{95\% CIs} & Estimate & \multicolumn{2}{c}{95\% CIs} & Estimate & \multicolumn{2}{c}{95\% CIs}  
      \tabularnewline
      \hline
      $\beta \times N$&0.99 & (0.07 & 1.25) & 0.9 & (0.67 & 1.14) & 1.23 & (1.03 & 1.42) \\ 
      $\gamma$&    0.14 & (0.1 & 0.16) & 0.12 & (0.1 & 0.14) & 0.15 & (0.13 & 0.17) \\ 
      $(\beta \times N)/\gamma$ &      7.19 & (0.65 & 9.17) & 7.5 & (5.98 & 9.03) & 8.27 & (6.8 & 9.88) \\ 
      $\alpha_0$ &     -6.28 & (-7.1 & -4.79) & -6.73 & (-7.58 & -5.68) & -6.59 & (-7.22 & -5.92) \\ 
      $\alpha_1$ &    -1.75 & (-2.26 & -1.13) & -1.96 & (-2.45 & -1.44) & -1.83 & (-2.28 & -1.38) \\ 
      $\alpha_2$  &    2.42 & (1.89 & 2.93) & 2.29 & (1.79 & 2.9) & 2.71 & (2.3 & 3.15) \\ 
      $\rho \times N$  &      68.5 & (40.9 & 83.4) & 65.7 & (53.2 & 80.4) & 79.2 & (65.7 & 95.4) \\ 
      \hline
    \end{tabular}
    \par\end{centering}
  \caption{Convergence diagnostics and sensitivity analysis: Posterior medians and 95\% equitailed credible intervals (CIs) under different initial values and assumptions for the parameters of the SIRS model estimated using clinical and environmental data sampled from Mathbaria, Bangladesh. The PMMH algorithm is run from different initial values using $N=10000$, $1/\mu=3$ years, $\phi_S/N=0.2$, and $\phi_I/N=0.02$, and also run using different values for the population size, $N$, the loss of immunity rate, $\mu$, and the means of the Poisson initial distributions, $\phi_S$ and $\phi_I$.}
  \label{RDests}
\end{table}

\clearpage
\renewcommand{\thefigure}{C-\arabic{figure}}
\setcounter{figure}{0}
\renewcommand{\thetable}{C-\arabic{table}}
\setcounter{table}{0}
\section*{Appendix C: Model fit}

\begin{figure}[h]
  \begin{centering}
    \includegraphics[width=0.85\linewidth]{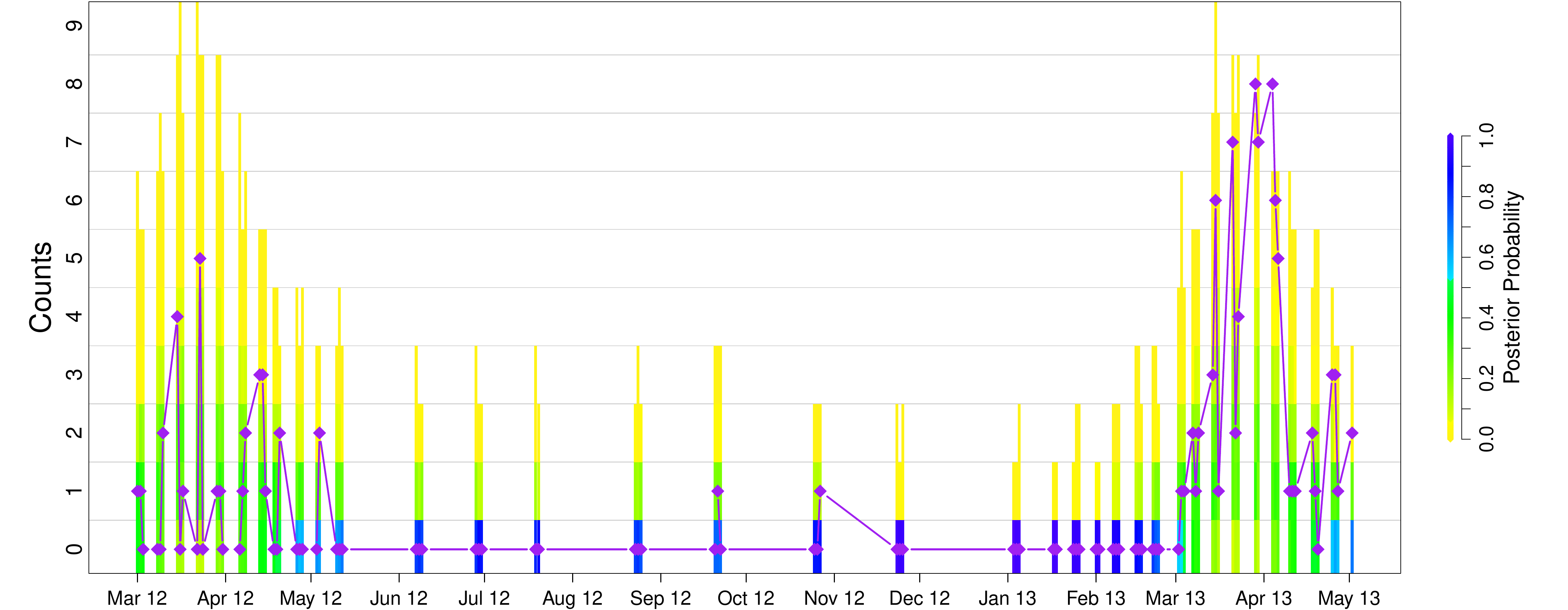}
    \includegraphics[width=0.85\linewidth]{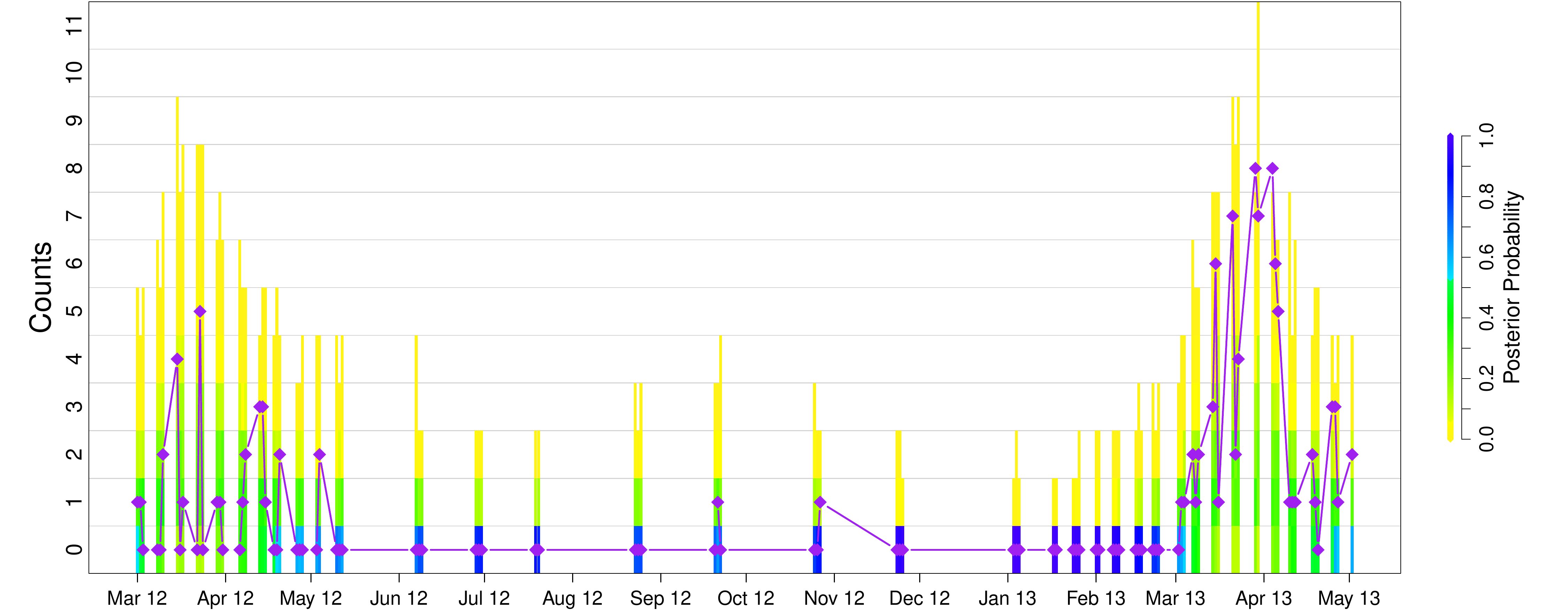}
    \includegraphics[width=0.85\linewidth]{ALLTIMESHeatMapWDWTLag21.pdf}
    \caption{Distributions of predicted reported cases under models
      assuming a covariate lag of 14 days (top), 18 days (middle), and
      21 days (bottom). The posterior probability of the predicted
      counts is compared to the test data (diamonds connected by
      straight line). The coloring of the bars is determined by the
      frequency of each set of counts in the predicted data for each
      time point. The distributions are similar, regardless of lag
      choice. }
    \label{kapHeat}
  \end{centering}
\end{figure}

\begin{figure}[h]
  \begin{centering}
    \includegraphics[width=0.85\linewidth]{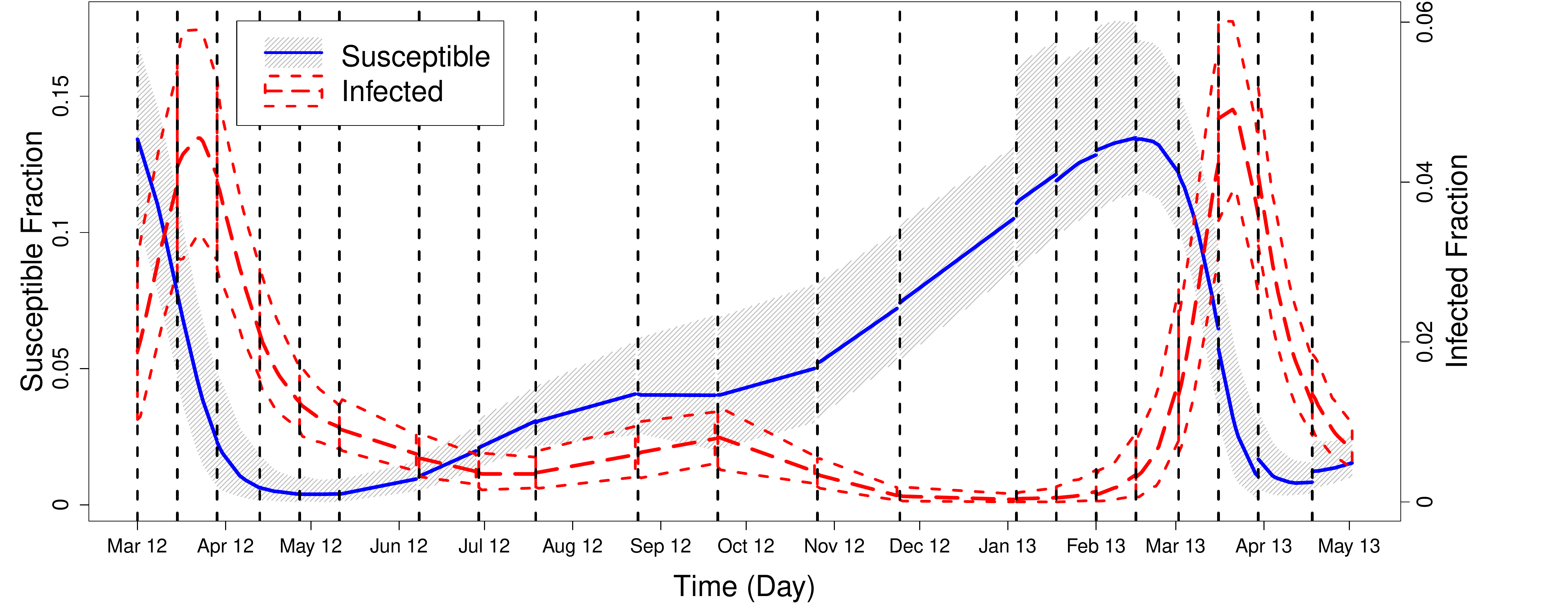}
    \includegraphics[width=0.85\linewidth]{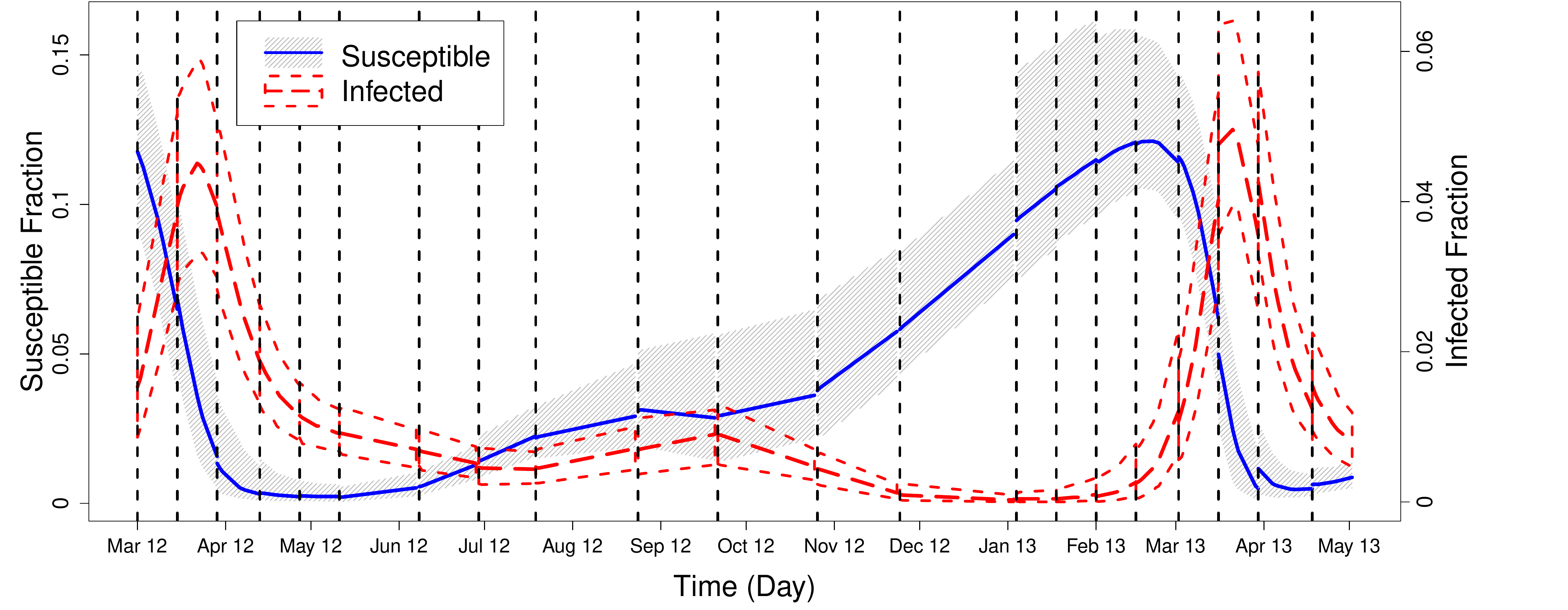}
    \includegraphics[width=0.85\linewidth]{ALLTIMEScombSUSandINFWDWTLag21.pdf}
    \caption{Predictive distributions of the hidden states, under
      models assuming a covariate lag of 14 days (top), 18 days
      (middle), and 21 days (bottom). The gray area and the solid line
      denote the 95\% quantiles and median of the predictive
      distributions for the fraction of susceptibles. The short dashed
      lines and the long dashed line denote the 95\% quantiles and
      median of the predictive distributions for the fraction of
      infected individuals. Differences between the distributions
      under the different lags are difficult to distinguish. }
    \label{kapHidden}
  \end{centering}
\end{figure}

To select a lag for the environmental covariates in the Mathbaria
analysis, we compare prediction results from models assuming three
different lags: $\kappa=14$, $\kappa=18$, and $\kappa=21$. These are
shown in Figures \ref{kapHeat} and \ref{kapHidden}. The predictive
distributions of the hidden states look similar across lags, so we use
the 21 day lag model in order to predict an upcoming epidemic furthest
in advance. With a three week lag, we would be able to make
predictions three weeks in advance.

\begin{figure}[h]
  \begin{centering}
    \includegraphics[width=1\linewidth]{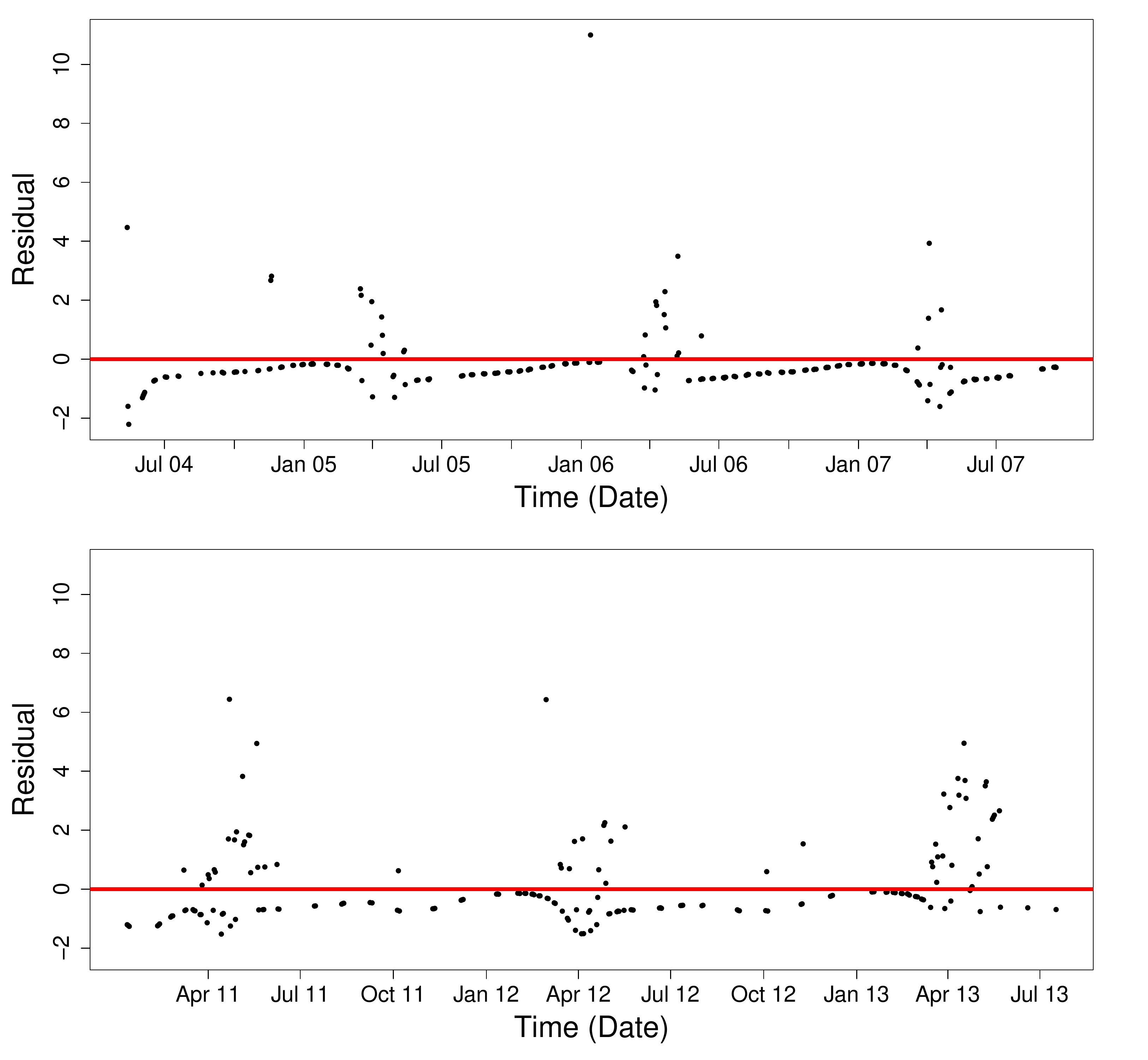}
    \caption{Plot of standardized residuals versus time. The top
      figure shows the residuals for the first three years of data
      collected from Bangladesh, and the bottom figure shows the
      residuals for the second three years of data collection. The red
      line is drawn through zero for reference. }
    \label{residuals}
  \end{centering}
\end{figure}

Figure \ref{residuals} shows plots of standardized residuals versus
time for each of the two phases of data collection in Mathbaria,
Bangladesh. Standardized residuals are calculated as
$\epsilon_{t_i}=\left(y_{t_i}-\mbox{E}(y_{t_i})\right)/\mbox{sd}(y_{t_i})$,
where $y_{t_i}$ is the number of observed infections at time $t_i$ for
observation $i \in \{0,1,...,n\}$. $\mbox{E}(y_{t_i})$ and
$\mbox{sd}(y_{t_i})$ are approximated via simulation by fixing the
model parameters to the posterior medians, running the SIRS model
forward 5000 times, and computing the average and sample standard
deviation of the 5000 realizations of the case counts at each time
point. Residuals are furthest from zero during the epidemic peaks; the
inflation of residuals during times of high case counts is probably
due to the model being off in terms of the timing of the epidemic peak
or the latent states not being predicted correctly.

\clearpage 
\renewcommand{\thefigure}{D-\arabic{figure}}
\setcounter{figure}{0} 
\renewcommand{\thetable}{D-\arabic{table}}
\setcounter{table}{0}
\section*{Appendix D: Sensitivity analysis}

In our analysis of the data from Mathbaria, we assumed the size of the
population, $N=10000$, the loss of immunity rate, $\mu=0.0009$, and
the means of the Poisson initial distributions, $\phi_S=.2 \times N$
and $\phi_I = 0.02 \times N$, are known. We studied sensitivity to
these assumptions by setting all of these parameters to different
values, and the results are shown in the bottom two-thirds of Table
\ref{RDests}. We report $\beta \times N$ and $\rho \times N$ since we
found these parameter estimates to be robust to changes in the
population size $N$. As seen in Table \ref{RDests}, estimates are
similar over different values of $N$, $\mu$, $\phi_S$ and $\phi_I$.

\begin{figure}[ht]
  \begin{centering}   
    \begin{tabular}{cc}
      \includegraphics[width=.5\linewidth]{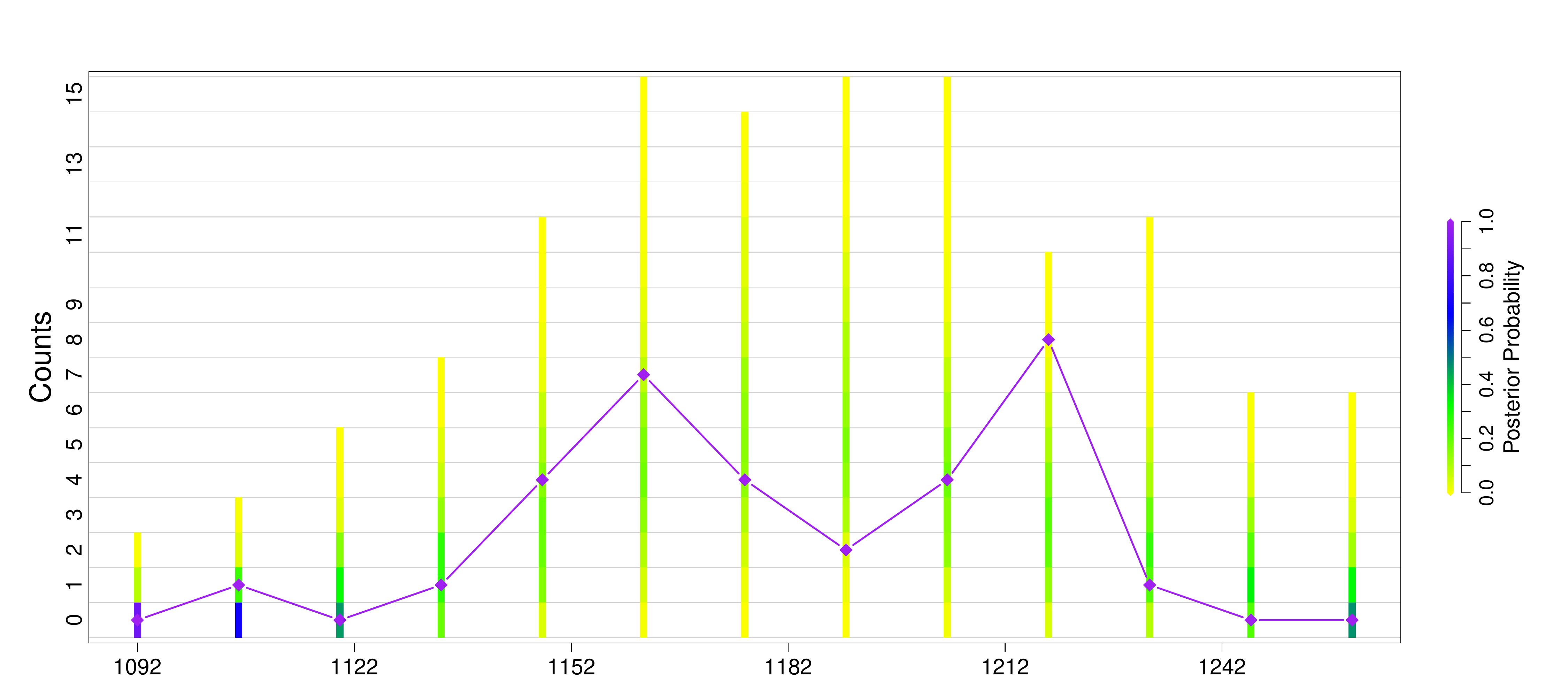} &    \includegraphics[width=.5\linewidth]{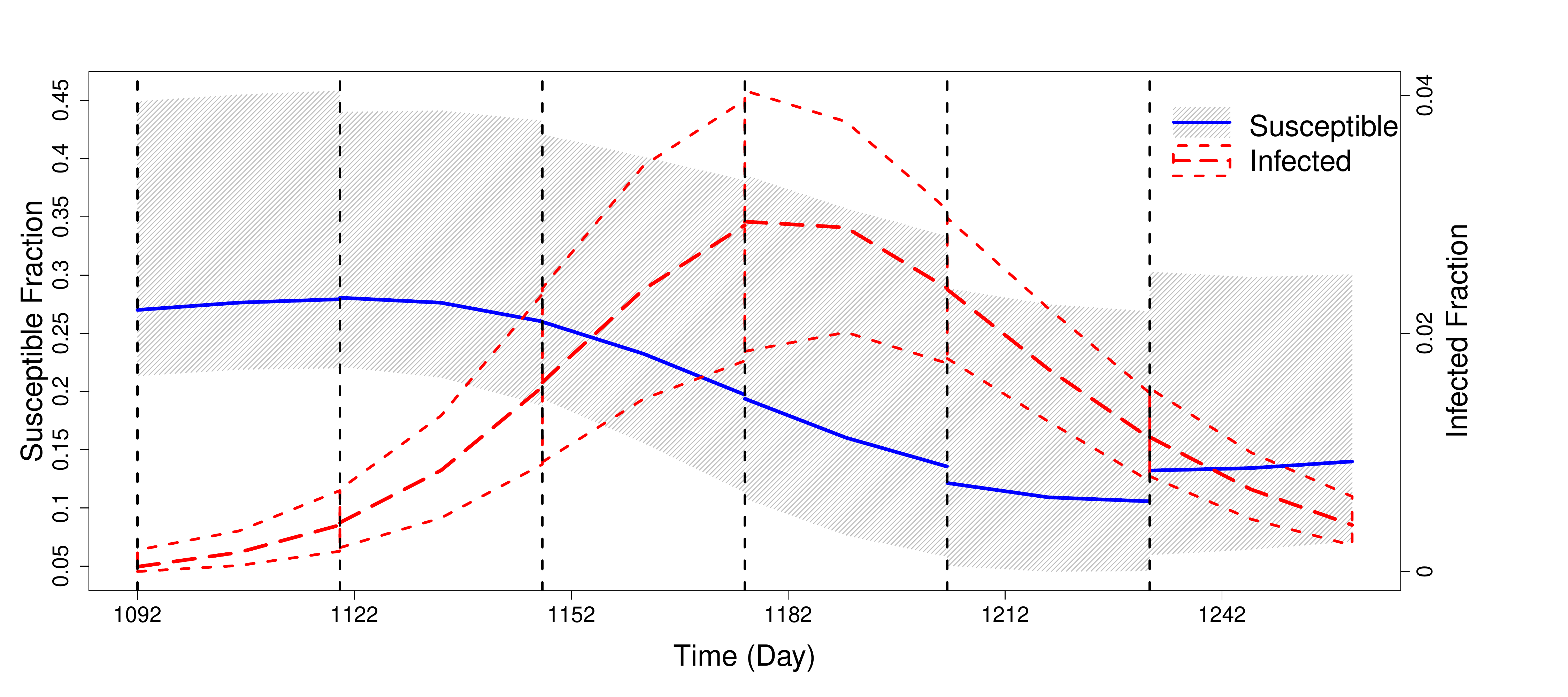}\\
      \includegraphics[width=.5\linewidth]{TRUTHALLTIMESHeatMap.pdf} &    \includegraphics[width=.5\linewidth]{TRUTHALLTIMEScombSUSandINF.pdf}\\
      \includegraphics[width=.5\linewidth]{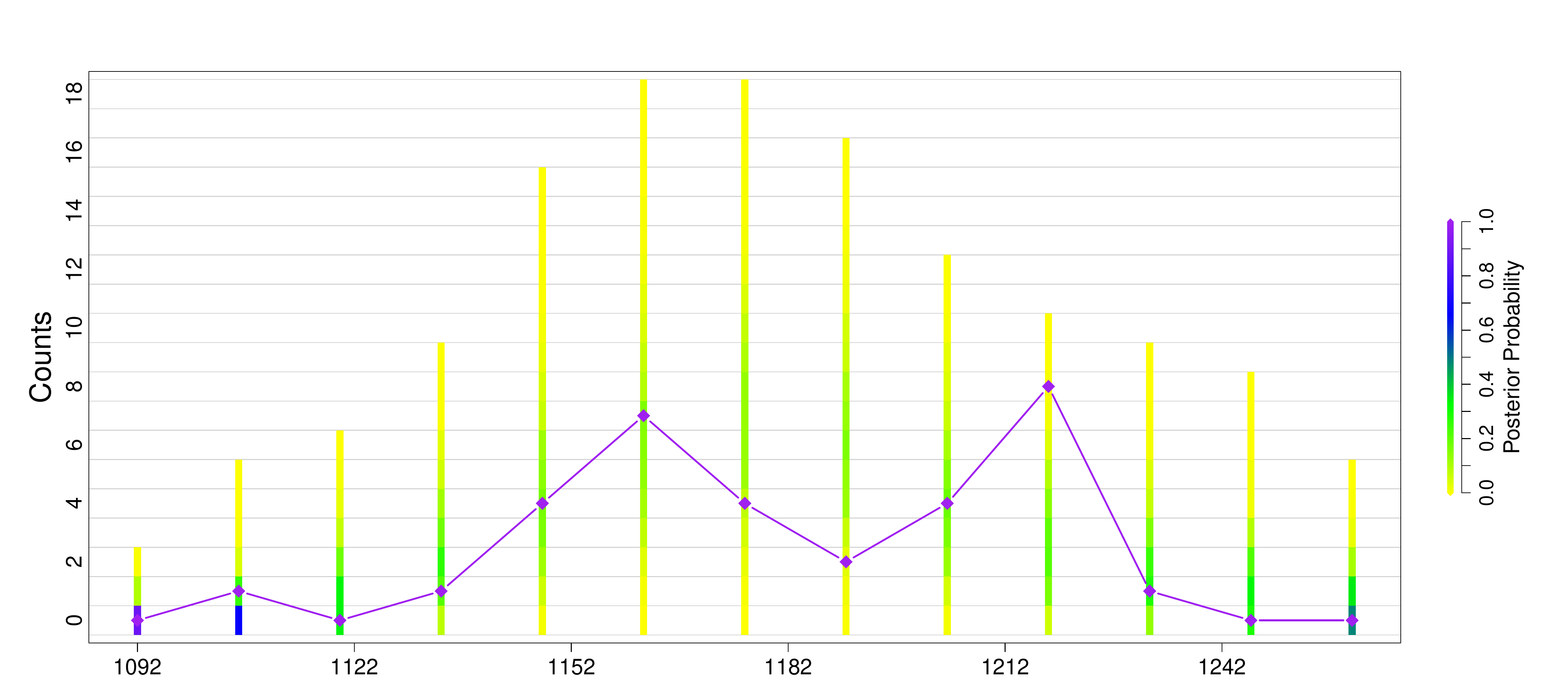} &    \includegraphics[width=.5\linewidth]{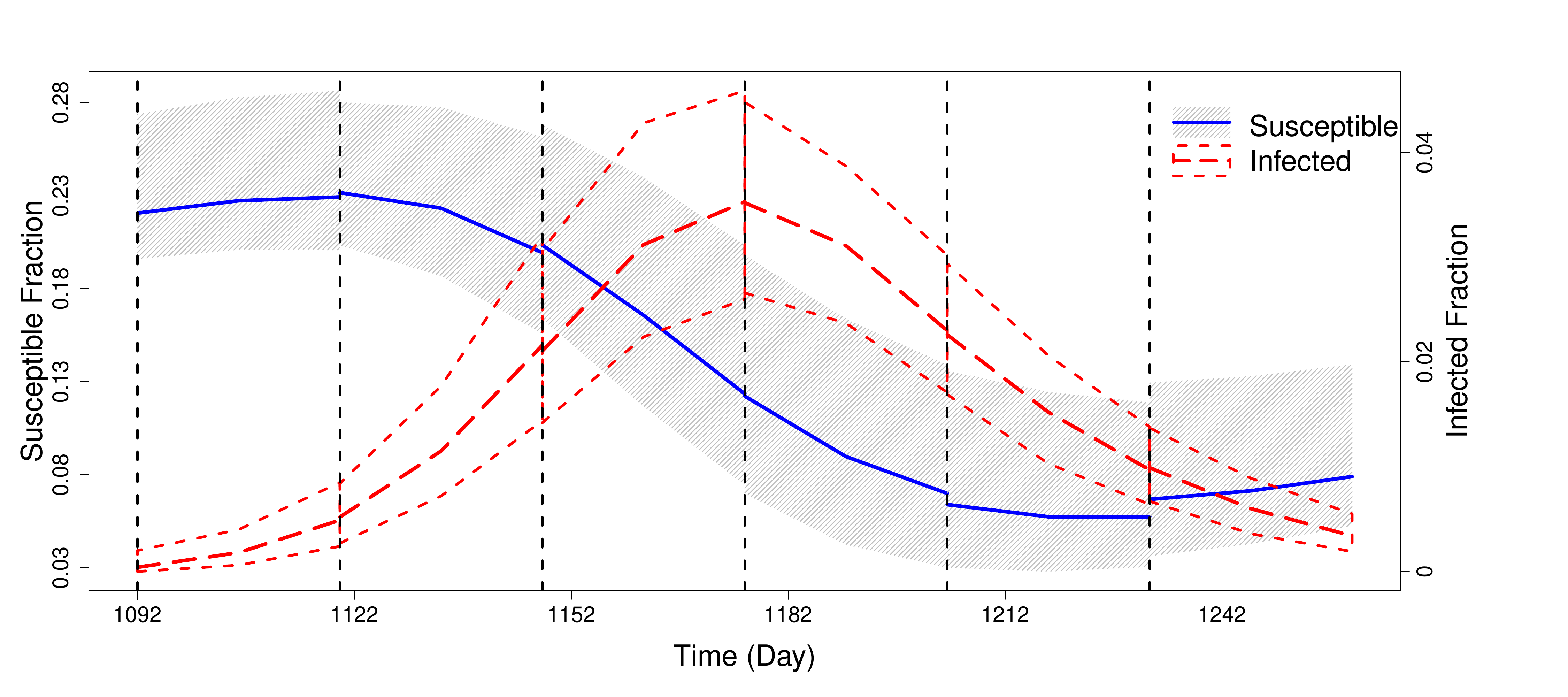}\\
      \includegraphics[width=.5\linewidth]{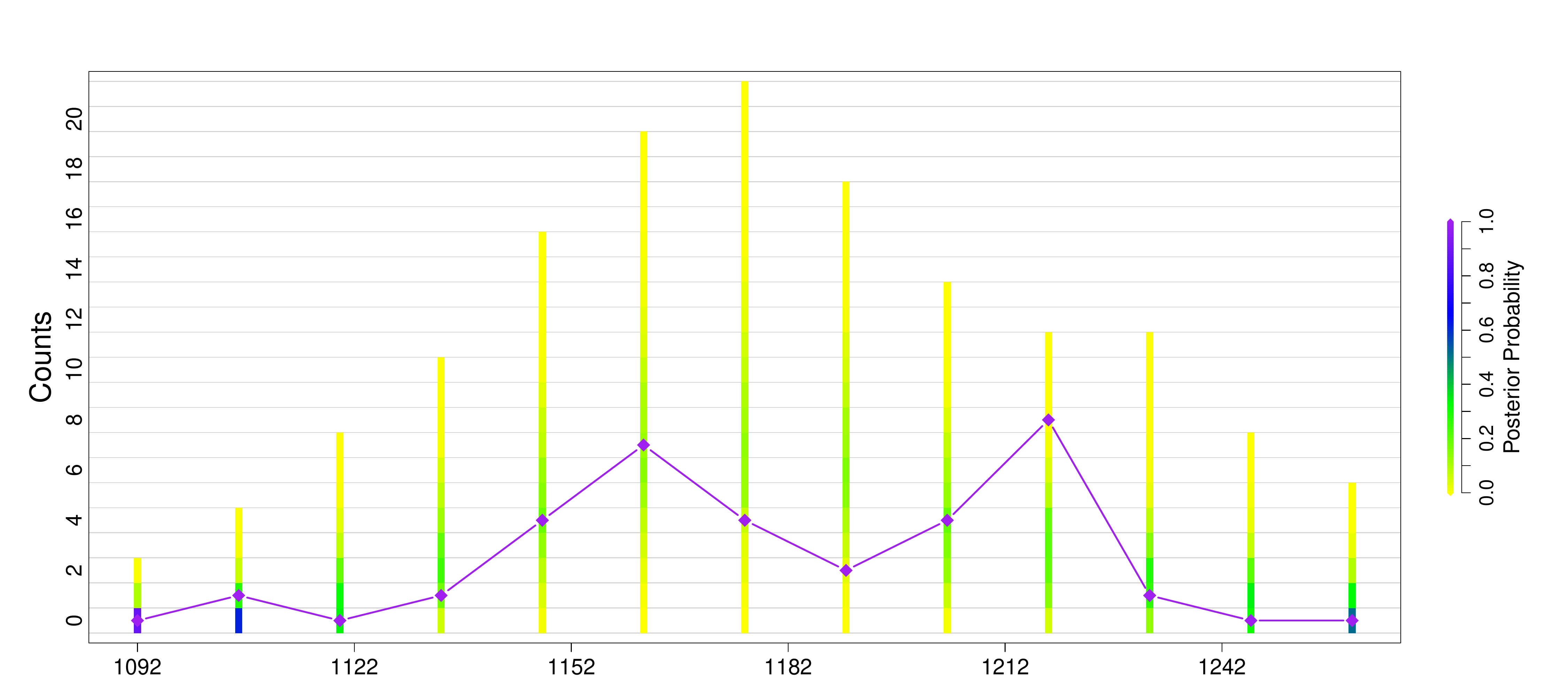} &    \includegraphics[width=.5\linewidth]{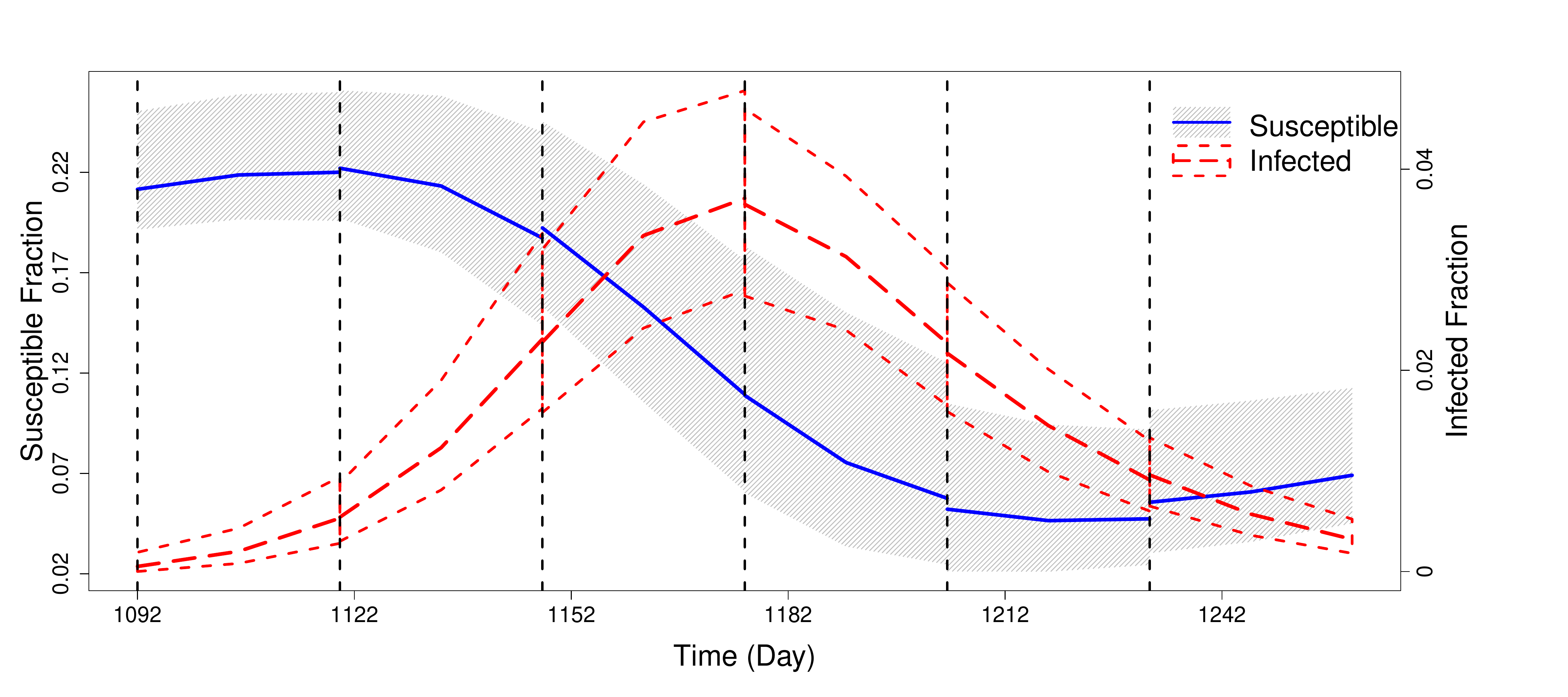}\\
    \end{tabular}
    \caption{Summary of prediction results for simulated data; each row
      shows prediction results under different assumptions about the
      values of $\phi_s/N$ and $\phi_I/N$. From top to bottom, values
      for $\phi_S/N$ and $\phi_I/N$ are set above the true values, (0.31,
      0.003), at the truth (0.21, 0.0015), under the true values (0.11,
      0.00075), and further under (0.055, 0.000375). Plots on the left
      compare the posterior probability of the predicted counts to the
      test data (diamonds connected by straight line). The coloring of
      the bars is determined by the frequency of each set of counts in
      the predicted data for each time point. The plots on the right
      show how the trajectory of the predicted hidden states change over
      the course of the epidemic. The gray area and the solid line
      denote the 95\% quantiles and median of the
      predictive distribution for the fraction of susceptibles. The
      short dashed lines and the long dashed line denote the 95\%
      quantiles and median of the predictive distribution
      for the fraction of infected individuals. 
    }
    \label{simphipred}
  \end{centering}
\end{figure}

We also tested the effect of incorrect values for $\phi_S$ and
$\phi_I$ on prediction using simulated data, as seen in Figure
\ref{simphipred}. Values for $\phi_S/N$ and $\phi_I/N$ are set above
the true values, (0.31, 0.003), at the truth (0.21, 0.0015), under the
true values (0.11, 0.00075), and further under the truth (0.055,
0.000375). Predicted distributions look similar for all values of
$\phi_S$ and $\phi_I$. Uncertainty is greatest when $\phi_S$ and
$\phi_I$ are set at higher values than the truth. For the lowest
values of $\phi_S$ and $\phi_I$, the fraction of susceptible
individuals is lower and the fraction of infected is higher than those
predicted fractions under other settings. However, important
information, like the timing of the epidemic, remains intact.

\clearpage 

\renewcommand{\thefigure}{E-\arabic{figure}}
\setcounter{figure}{0}
\renewcommand{\thetable}{E-\arabic{table}}
\setcounter{table}{0}
\section*{Appendix E: Prediction}

\begin{figure}[h]
  \begin{centering}
    \includegraphics[width=1\linewidth]{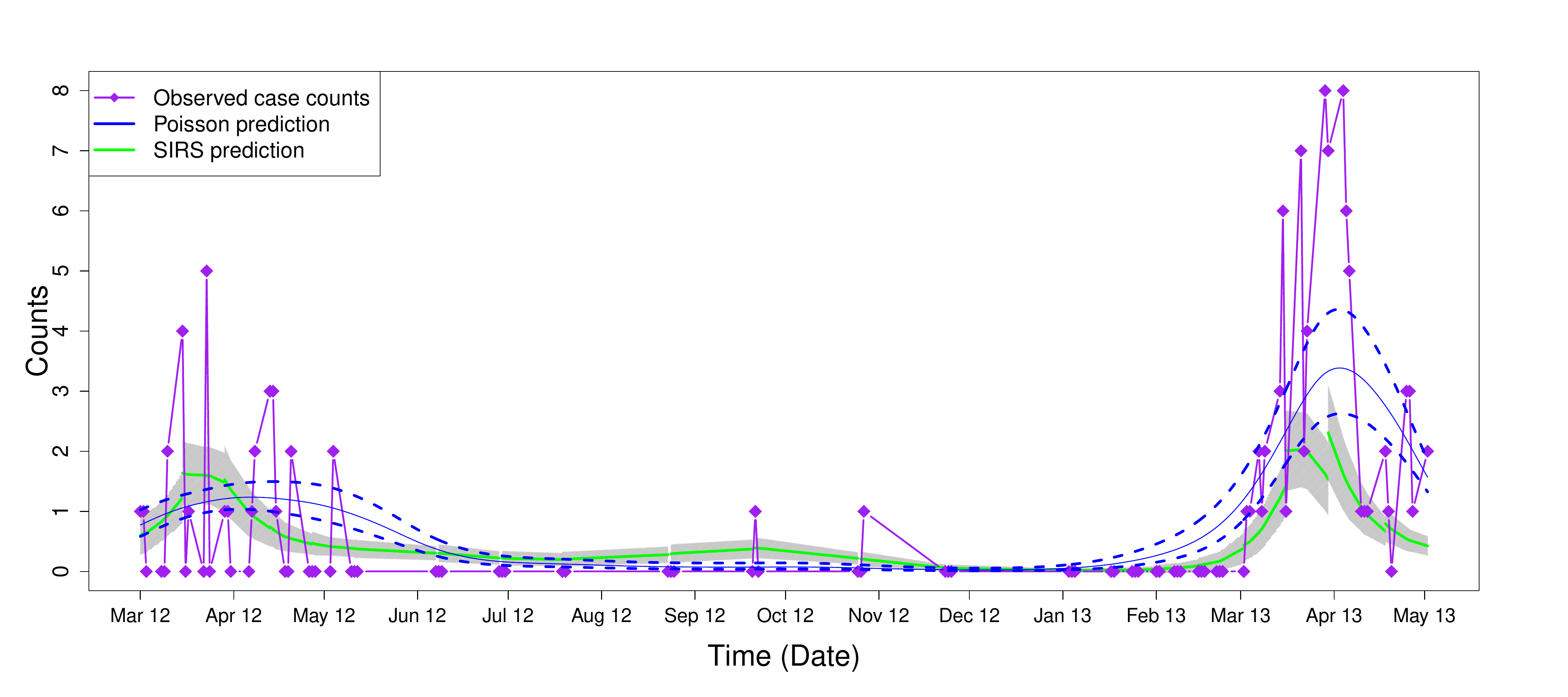}
    \caption{Comparison of predicted means for number of reported
      cases. The solid blue lines and the dashed blue lines denote the
      predicted means and 95\% intervals under the
      quasi-Poisson model. The green line and gray area denote the predicted means and 95\% intervals under the SIRS model. Predictions are
      started and stopped using identical cut-off points for the
      training and test data to those in Figure \ref{mathpred}. Test
      data are denoted by the purple diamonds connected by straight
      lines.
    }
    \label{poispred}
  \end{centering}
\end{figure}

We compare SIRS predictive distributions to predictions made from a
lagged quasi-Poisson regression model, similar to the one used by
\cite{phase1}. For the two predictors, water temperature (WT) and
water depth (WD), we have
\[
\mbox{ln}\mbox{ E}(Y_t|C_{WD}(t-\kappa),C_{WT}(t-\kappa))
=\beta_0+\beta_1C_{WD}(t-\kappa)+\beta_2C_{WT}(t-\kappa),
\]
where $\kappa=21$ days. The quasi-Poisson model accounts for
overdispersion in the data \citep{GLMs}. Figure \ref{poispred} shows
the predicted means and 95\% intervals under the quasi-Poisson
model. Test data are again cut off at different points during the 2012
and 2013 epidemic peaks and predictions are run until the next cut off
point, with cut off points chosen approximately every two weeks.
Predicted mean number of reported cases and 95\% intervals from the
hidden SIRS model are also shown for comparison. To calculate these,
we sample 500 sets of parameter values from the posterior. For each
set of parameters, we simulate data forward until the next cut off
point 100 times and then calculate the mean of the predicted counts at
each observation time. Using these 500 means from the 500 parameter
sets, we calculate the overall predicted means and 95\%
intervals. Both models predict well the timing of epidemic
peaks. However, the quasi-Poisson regression framework does not
provide any information about the underlying fraction of infected
individuals in the population, which may be important for resource
allocation.

\clearpage 

\renewcommand{\thefigure}{F-\arabic{figure}}
\setcounter{figure}{0}
\renewcommand{\thetable}{F-\arabic{table}}
\setcounter{table}{0}
\section*{Appendix F: Routes of transmission}

\begin{figure}[h]
  \begin{centering}
    \includegraphics[width=.7\linewidth]{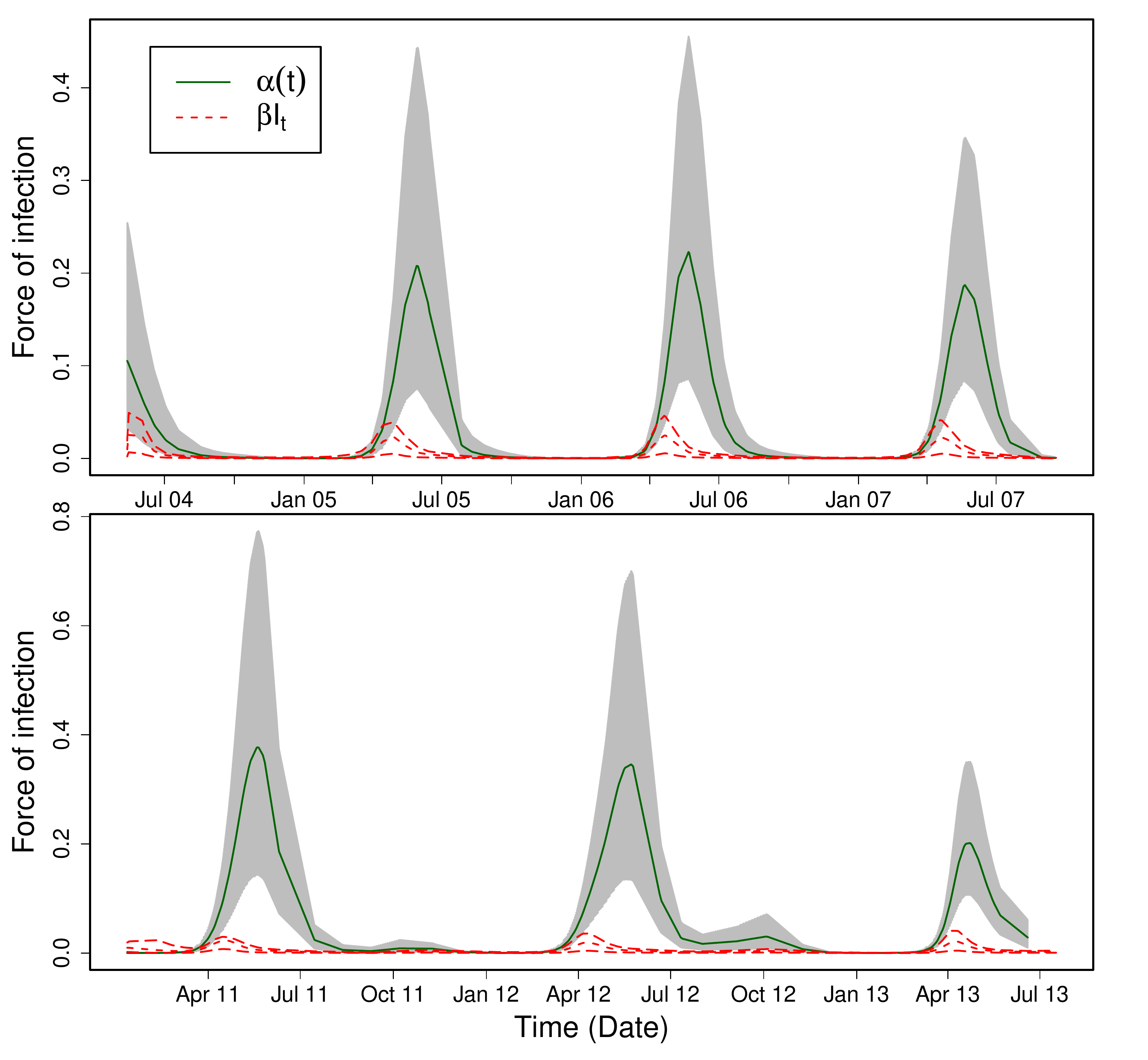}
    \caption{The relative contribution of different routes of
      transmission at different points of the epidemic curves. The
      gray area and the solid line denote the 95\% quantiles and
      median of the force of infection from the environment,
      $\alpha(t)$. The long dashed lines and the short dashed line
      denote the 95\% quantiles and median of the force of infection
      from infected individuals, $\beta \times I_t$.}
    \label{alphavbetaI}
  \end{centering}
\end{figure}

An important question in cholera modeling is: what is the relative
contribution of different routes of transmission at different points
of the epidemic? We hypothesized that environmental forces trigger the
seasonal cholera epidemics and that infectious contact between
susceptible and infected individuals drives the epidemics. To examine
this possible dynamic, we compare the forces of infection from the
environment, $\alpha(t)$, to that from infected individuals, $\beta
\times I_t$, over time. Values are computed by sampling 5000 sets of
parameter values from the posterior. For each set of parameters, we
generate data using our hidden SIRS model. Figure \ref{alphavbetaI}
shows median and 95\% quantiles for $\alpha(t)$ vs $\beta \times I_t$
plotted over time. The median values of $\alpha(t)$ are almost always
higher than values of $\beta \times I_t$; the only time it is not (in
early 2011) is most likely caused by model misspecification,
specifically in setting $\phi_I$ to the wrong value for this
phase. This model misspecification is similarly reflected in Figure
\ref{residuals}. This supports the hypothesis that the epidemics are
driven by the environmental force of infection.  \vspace{-10pt}

\clearpage

\bibliographystyle{plainnat}
\bibliography{pmcmc}{}

\end{document}